\documentclass[ aps,pra,twocolumn, superscriptaddress, amsmath,notitlepage,
amssymb ]{revtex4-1}

\usepackage[utf8]{inputenc}
\usepackage{graphicx}
\usepackage[colorlinks=true,hypertexnames=false]{hyperref}
\usepackage{amsmath}
\usepackage{amssymb}
\usepackage{braket}
\usepackage{bbold}
\usepackage{times}
\usepackage{color}

\DeclareMathOperator\artanh{artanh} \newcommand{\eq}[2]{\begin{equation} #1
\label{eq:#2}\end{equation}} 
 
\newcommand{\pf}{\mathrm{Pf}} \newcommand{\dd}{\mathrm{d}}
\newcommand{\id}{\mathbb 1}

\begin{document}

 \title{Topological dualities via tensor networks}

\author{C.~Wille} 
\affiliation{Rudolf Peierls Centre for Theoretical Physics, Oxford OX1 3PU, UK}

\author{J.~Eisert} 

\affiliation{Dahlem Center for Complex Quantum Systems, Freie Universit{\"a}t
Berlin, 14195 Berlin, Germany} \affiliation{Helmholtz-Zentrum Berlin f{\"u}r
Materialien und Energie, 14109 Berlin, Germany}

\author{A.~Altland}
\affiliation{Institut f\"ur Theoretische Physik, 50937 Cologne, Germany}

\begin{abstract}
The ground state of the toric code, that of the two-dimensional class D superconductor, and the partition sum of the two-dimensional Ising model are dual to each other. This duality is remarkable inasmuch as it connects systems commonly associated to different areas of physics -- that of long range entangled topological order, (topological) band insulators, and classical statistical mechanics, respectively. Connecting fermionic and bosonic systems, the duality construction is intrinsically non-local, a complication that has been addressed in a plethora of different approaches, including dimensional reduction to one dimension, conformal field theory methods, and operator algebra. In this work, we propose a unified approach to this duality, whose main protagonist is a tensor network (TN) assuming the role of an intermediate translator. Introducing a fourth node into the net of dualities offers several advantages: the formulation is integrative in that all links of the duality are treated on an equal footing, (unlike in field theoretical approaches) it is formulated with lattice precision, a feature that becomes key in the mapping of correlation functions, and their possible numerical implementation. Finally, the passage from bosons to fermions is formulated entirely within the two-dimensional TN
framework where it assumes an intuitive and technically convenient form.  We illustrate the predictive potential of the formalism  by exploring the fate of phase transitions, point and line defects,  topological boundary modes, and other structures under the mapping between system classes.  Having
condensed matter readerships in mind, we  introduce the construction pedagogically in a manner assuming only minimal  familiarity with the concept of TNs.
\end{abstract}
\date{\today}

\maketitle

\section{Introduction}

Where they exist, dualities are powerful aides in understanding the physics of
nominally different complex systems. As a case in point, consider the ground
state of the \emph{toric code} (TC), that of a topological \emph{superconductor} (SC) in symmetry class D, and the partition sum of the classical two-dimensional \emph{Ising model}
(IM) --- three of the main protagonists of this work. In a sense to be made  precise in the following, these
systems are connected by duality transformations \cite{Blackman,PhysRevB.44.4374,PhysRevE.58.1502,Berezin_1969,PlechkoEnglish,Dotsenko83,ChalkerMerz,Castelnovo,Zhu}. In the case at hand, these draw connections
between bosonic and fermionic systems, ground states and partition sums, and
between classical and quantum systems. They also link  systems which
are at the forefront of interest to different communities. For example, the
toric code ground state is a paradigmatic example of a long range entangled 
state of matter (hence featuring intrinsic topological order) \cite{Kitaev-AnnPhys-2003}, while the topological superconductor is a free fermion system belonging to the family of topological insulators \cite{Ryu_2010}.
All three systems display phase transitions --- between an ordered phase and a
topological spin liquid, a trivial and a topological superconductor and a ferro-
and a paramagnet, respectively --- and the duality establishes the equivalence
between these. The same applies to physics at various defect structures, for
example, the formation of gapless boundary modes in the superconductor related
to the behavior of anyonic excitations at the boundary of the toric code.

Dualities in condensed matter physics are generically established via a toolbox
of recurrent concepts. These include the mappings between $d$-dimensional
quantum systems and $(d+1)$-dimensional partition sums, the taking of continuum
limits mapping to (conformal) field theories and dimensional analysis, or the
comparison of operator commutator algebras on different sides of the duality.
For example, one way to go from the two-dimensional Ising model to the
superconductor, is to first apply an anisotropic scaling deformation to map the
former to the transverse magnetic field quantum Hamiltonian, then equate this
bosonic system to a fermionic Majorana chain via Jordan-Wigner
\cite{Onsager,Kaufman,SCHULTZ} transformation, and finally re-discretize time to
arrive at the two-dimensional lattice Hamiltonian describing a superconductor in
the Majorana basis \cite{Dotsenko83,PlechkoEnglish,ChalkerMerz}.

In this work, we consider the TC/SC/IM triplet to illustrate how \emph{tensor
networks} (TN) offer an efficient and intuitive alternative approach to duality\cite{Vanhove_2018,aasen2020topological,Lootens2023,lootens2022dualities}. 
The idea is to place a TN in the so-called matchgate category~\cite{Valiant,Bravyi2009,Jahn,Cai,Cai2007OnTT} as an intermediate
between the three systems. There are manifold advantages to bringing in a fourth
system as a translator. First, the TN comes in two incarnations, a bosonic and a
fermionic one, and the passage between the two is established directly on the
two-dimensional lattice by what in effect is a `two-dimensional Jordan-Wigner
transformation' \cite{Onsager,Kaufman,SCHULTZ,PhysRevB.104.035118}. In this way,
we may pass from bosons to fermions avoiding dimensional detours. (The operation
is conceptually similar, but somewhat more direct than previous constructions
\cite{Berezin_1969,Dotsenko83} based on the commutator algebra.) 

Second, the mapping is microscopic and explicitly relates to the operator 
contents of all three theories. This level of
detail, which is lost in continuum approaches,  supports intuition and is essential in the construction of dual representations of 
correlation functions. We will illustrate this point on the equivalence between free Majorana correlation functions of the SC with more complex correlations between composites of spin and disorder operators \cite{KadanoffCeva,Fradkin_2017}
(for a definition of disorder operators, 
see below)
in the IM.  

Finally, the approach keeps all three partners of the duality construction in
permanent sight. In this regard, it is different from previous approaches
focusing on one specific link of the duality. The principle behind this high
level of versatility is that a tensor network per se (unlike a Hamiltonian) has
no pre-assigned physical interpretation. More precisely, while for a TN with
\emph{open} indices these are identified with physical degrees of freedom, for a
TN without open indices, the interpretation of individual tensors and their
indices  is not canonical. This ambiguity can be exploited to yield relations
between seemingly unrelated physical systems. For example, the partition sum of
a two-dimensional statistical mechanics model with local interactions affords
interpretation in terms of  a two-dimensional tensor network. Alternatively,
consider the ground state of a spin system  represented as a \emph{projected
entangled pair state} (PEPS), i.e., a tensor network on a two-dimensional graph
with open indices corresponding to the Hilbert spaces of the local spins
\cite{Cirac21}. The overlap of this state with itself can again be interpreted
as a classical partition function. It is then natural to go one step further and
establish a \emph{local} correspondence between these systems by looking at the
respective fine structure of their TN representations.

Supplemented with a boson-fermion mapping on the level of the TN, these ideas
become even more powerful. Below, we will use such ambiguities of TN
representations as a resource to discuss  the full web of dualities in a
comprehensive manner. In particular, we discuss a local equivalence between the
partition sum of the IM and expectation values of the ground state of the toric
code with string tension \cite{Castelnovo}. After a boson-fermion mapping, one can furthermore
equate the TN to a Grassmann integral describing the ground state of a
band-insulator Hamiltonian in symmetry class D.

While the technical elements of this construction are known in the theory of
matchgate tensor networks, we here present them in a comprehensive manner,
aiming to introduce the key ideas to the community of condensed matter
physicists. This endeavor is not just of pedagogical value: all three systems
linked by the duality show rich behavior when translational symmetry is broken
via the introduction of spatial phase boundaries, or defect structures. Examples
include vortices, and the formation of gapless boundary modes in the
superconductor, the binding of anyonic excitations at the ends of line defects
in the toric code, or the physics carried by string-like `disorder operators' in
the Ising model. All these are  subject of the duality mapping.
However, the specific ways in which they transform are not always obvious. For
example,  the above-mentioned Majorana correlation function  probes the propagation of quasi-particles injected into the superconductor ground state from one point to another. In the Ising context it becomes more complex, and now describes the correlations
of a composite of a spin- and a disorder operator. The definition of the latter
is non-trivial because it responds to the precise positioning of the composite
operator on the Ising lattice \cite{KadanoffCeva,Fradkin_2017}. 

This and various other mappings will illustrate the application of the TN
construction and are meant to introduce some powerful tricks of TN algebra to
condensed matter practitioners. In a follow up publication \cite{Preparation} we
will push this framework into less charted territory, including the presence of
translationally invariance breaking disorder, the inclusion of nonlinear
correlations in the TN, and that of geometrically distorted (`holographic')
background geometries \cite{Preparation2}.

The remainder of this work is structured as follows. In Section~\ref{sec:MG}, we
introduce the formalism of matchgate tensor networks and their representation as
free fermion partition functions. Section~\ref{sec:dualities} is the main focus
of this work. It contains the derivation of the aforementioned dualities for the
translation invariant case and discusses the phase transitions across the
different systems. In Section~\ref{sec:inhomogeneous_structures}, we extend the
dualities to situations where translation invariance is broken and discuss how
the duality map between different correlation functions. We summarize our work
in Section~\ref{sec:sum} and provide an outlook to future research.

\section{Matchgate tensor networks}\label{sec:MG} The workhorse by which the
connections discussed in this work will be drawn are \emph{matchgate tensor
networks} (MGTN) \cite{Valiant,Bravyi2009,Jahn,Cai,Cai2007OnTT}. In the
following, we introduce these objects in a manner assuming only a minimal level
of familiarity with TNs (for introductory reviews,  see
Refs.~\cite{Orus-AnnPhys-2014,AreaReview,Bridgeman2017,Orus2019,Cirac21}). 

\subsection{Bosonic matchgate tensors}

\paragraph*{Matchgate tensors.}
A bosonic normalized, even \emph{matchgate} (MG) tensor $T$ tensor  carries  $n$
indices $i_j=0,1$ of bond dimension two. As such it is described by complex
coefficients $T_{i_1 i_2\ldots i_n}$ such as $T_{0100\dots}$. To make it a
matchgate tensor, we need to add further structures. Concretely, a matchgate tensor satisfies the following rules.
\begin{itemize}
\item (Normalization) $T_{0\ldots 0}=1$.
\item (Evenness)	 If $i_1+\ldots+i_n$ is odd, $T_{i_1\ldots i_n}=0$.
\item (Gaussianity) The ${n \choose 2}=n(n-1)/2$ so-called second moments
$T_{i_1 \ldots i_n}$ with $i_1+\ldots+i_n=2$ are independent. We collect them in
an anti-symmetric $n\times n$- matrix $A^T=-A$, where $A_{12}=T_{110\ldots0}$,
$A_{13}=T_{1010\ldots 0}$, etc.
\end{itemize}
By definition, higher moments of $T$, i.e., entries with $i_1+\ldots + i_n>2$,
are given by \emph{Pfaffians} of submatrices of $A$. These submatrices are
obtained by deleting all rows and columns $k$ for which $i_k=0$. For example,
for a tensor with six indices the tensor entry $T_{111100}$ is given by $\pf
A|_{56}$, where $A_{56}$ is the sub-matrix of $A$ obtained by deleting the rows
and columns 5 and 6. For a tensor with four indices, the matrix  
\begin{equation} \label{eq:A}
	A=\begin{pmatrix}
		0& a_{12} & a_{13} & a_{14} \\
		-a_{12} &0 &a_{23} & a_{24} \\
		-a_{13} & - a_{23} &0 &a_{34} \\
		-a_{14} & -a_{24} & -a_{34} &0 
	\end{pmatrix}
\end{equation}
uniquely specifies all tensor entries. For example, $T_{1100}=a_{12}$,
$T_{1010}=a_{13}$, etc. The only non-trivial higher moment is given by
$T_{1111}=\pf(A)$.

\begin{figure}
	\centering
	\includegraphics[width=0.22\textwidth]{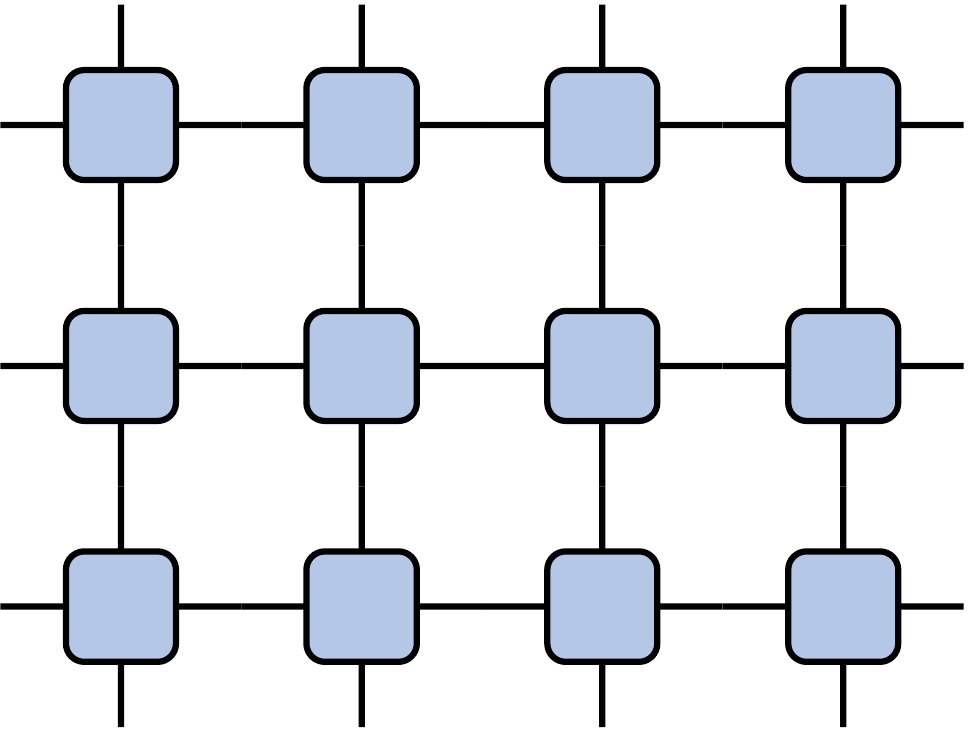}
	\hfill
	\includegraphics[width=0.22\textwidth]{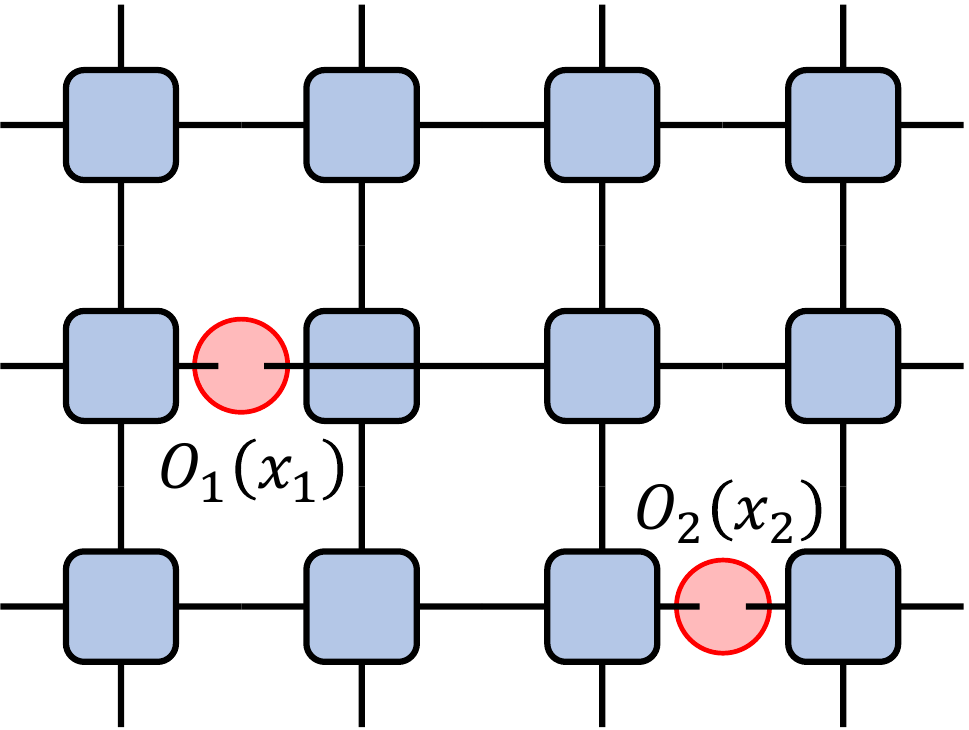}
	\caption{Translation invariant two-dimensional square lattice tensor network
	of tensors $T$ (left) and tensor network with incisions (right) that allow
	to calculate arbitrary $n$-point correlation functions of local observables,
	here $\langle O_1(x_1) O_2(x_2) \rangle$.}
	\label{fig:tn}
\end{figure}

\paragraph*{Matchgate tensor networks.}
We now consider two-dimensional square-lattice networks of matchgate tensors as
shown in Fig.~\ref{fig:tn}. First consider the fully contracted TN, without
open physical indices. As such, it is just a number (much as a partition sum is
just a number), and not very interesting in its own right. To obtain information about, e.g., correlations, we may 
cut bonds to define  open indices. For example we can calculate
the  correlation function $\langle O_1(x_1) O_2(x_2)\rangle$ of local
observables $O_{1,2}$ by incisions at two points $x_1,x_2$ as shown in
Fig.~\ref{fig:tn}.

\subsection{Mapping to Gaussian   fermionic tensor networks}

Bosonic matchgate tensor networks afford a reinterpretation as Gaussian
fermionic tensor networks~\cite{Bravyi2009}. To establish this connection, we
first review the concept of (Gaussian) fermionic tensor networks as such.

\emph{Fermionic tensor networks}
have been introduced to represent  many-body  states of fermions
\cite{PhysRevA.80.042333,CorbozPEPSFermions,PhysRevB.81.245110,Kraus,gu2010grassmann,Nick}.
In one~\cite{gu2010grassmann} of several optional representations a fermionic
tensor $T_\text{f}$ with $n$ indices of bond dimension two contains $n$
Grassmann variables $\theta_j$
\begin{equation} T_\text{f} = T_{i_1 \ldots i_n} \theta_{1}^{i_1} \ldots
\theta_{n}^{i_n} \;
\end{equation}
(see also Refs.~\cite{PhysRevB.95.245127}). We may identify these with fermion
 states, $\theta_i \mapsto c_i^\dagger |0\rangle  $, which are either occupied
 or unoccupied depending on the value $i_j=0,1$. Throughout, we work with
 tensors of even fermion parity, $T_{i_1\ldots i_n}=0$ if $i_1 + \ldots + i_n
 \mod 2 =1$. 

For two tensors $A=A_{i_1 i_2 \ldots}\theta_{A1}^{i_1} \theta_{A2}^{i_2} \ldots$
and $B=B_{k_1 k_2 \ldots }\theta_{B1}^{k_1}  \theta_{B2}^{k_2} \ldots$, the
formal product $A B$ represents a superposition of  states with up to $2n$
fermions via the above identification. We define a \emph{contraction} of indices
$Aj$ and $Bl$    
 by projection onto all states with equal occupation of $Aj$ and $Bl$ fermions.
The basic  mathematical identity realizing this contraction reads
\begin{align*}
	\int \dd \theta_{Aj} \dd \theta_{Bl}\, e^{\theta_{Bl} \theta_{Aj}} \theta_{Bl}^{k_l}\theta_{Aj}^{i_j} =\delta^{k_l i_j}.
\end{align*}
This contraction carries an orientation as contracting $Aj$ with $Bl$ differs
from contracting $Bl$ with $Aj$. Before using the identity above the Grassmann
variables $\theta_{Aj}$ and $\theta_{Bl}$ first need to be permuted through the
remaining variables such that they come to stand upfront in the indicated order.
The contraction of generic indices thus comes with a sign factor.

The generalization to multiple tensors $T^\alpha$, $\alpha=1, \dots ,N$ with
$n_\alpha$ fermions each is straightforward: introduce the vector $\underline
\theta = (\theta_{11},\ldots, \theta_{1n_1},\ldots, \theta_{N1},\ldots
\theta_{Nn_N})$ containing all  fermionic modes, and an anti-symmetric matrix
$C$ indicating the pattern of (oriented) contractions as  $C_{\alpha i,\beta j}
= 1$, if the $i$-th mode of tensor $T^\alpha$ is contracted with the $j$-th mode
of tensor $T^\beta$. The contraction of the network is then implemented by the
integral
\begin{equation}
\text{TN}_{(C,T)}=	\int (\dd \theta)_C \,e^{\frac 1 2 \underline \theta^T C \underline \theta} \; T^1  \ldots T^N \;, \label{eq:contracted}
\end{equation}
where $(\dd \theta)_C$ is a shorthand notation for the product of all ordered
pairs $\dd \theta_{\alpha i} \dd \theta_{\beta j}$ with $C_{\alpha i,\beta
j}=1$.

\emph{Gaussian  fermionic tensor networks.} A tensor $T$ with $n$ indices is a
\emph{fermionic Gaussian} (fG) tensor  if there exists a real anti-symmetric $n
\times n$-matrix $A=-A^T$ such that
\begin{equation}
	T_\text{fG} = e^{\frac 1 2 \underline \theta^T A \underline \theta} \;,\quad \underline \theta^T=(\theta_1,\ldots,\theta_n)\;. \label{eq:fG}
\end{equation}
The tensor product of two fG tensors $T_1$, $T_2$ is again a Gaussian tensor
given by
\begin{equation}
	T_\text{1}  T_\text{2} = e^{\frac 1 2 \underline \theta^T (A_1 \oplus A_2) \underline \theta} \;, \quad \underline \theta=(\underline \theta_1,\underline \theta_2)\;.
\end{equation}
Including the contractions in Eq.~\eqref{eq:contracted}, we write the contracted
fermionic Gaussian tensor network as
\begin{equation}
	\label{eq:TfGContraction}
	\text{TN}_{(C,A)}=\int (\dd \theta)_C \,e^{\frac 1 2 \underline \theta^T (A+C) \underline \theta} \;, 
\end{equation}
where $A=\oplus_i A_i$ is the direct sum of all individual characteristic
functions of the tensors $T_i$. 

Note that a  real fermionic Gaussian tensor network can be interpreted as the
partition sum $Z=\int \dd \underline \theta e^{-S}$ with the weight
\begin{equation}
	S= \frac {\mathrm i} 2  \theta^T  H \theta \;, \quad H=\mathrm i (A+C)\;, \label{eq:HCA}
\end{equation}
where $H=H^\dag$ and $H=-H^T$. Within the framework of the tenfold symmetry
classification of free fermions systems, this is a Hamiltonian in symmetry class
D.

\emph{Matchgates as Gaussian fermionic tensors.}
Eq.~\eqref{eq:fG} implies the advertised  connection between  fermionic Gaussian
and bosonic matchgate tensors: Given a bosonic  $T_\text{MG}$ with second
moments $A$, we define  
\begin{align}
	\label{eq:TMGtoTfG}
	T_\text{fG}= (T_\text{MG})_{i_1\ldots i_n}  \theta_{1}^{i_1} \ldots \theta_{n}^{i_n}  = e^{\frac 1 2 \underline \theta^T A \underline \theta} \;.
\end{align}
The indicated ordering of Grassmann variables is an essential element of the map
$T_\textrm{MG}\mapsto T_\textrm{fG}$.

The assignment Eq.~\eqref{eq:TMGtoTfG} remains  formal unless we have settled
the following consistency issue: For a (partially) contracted matchgate tensor
network one may either first turn to a fermionic representation of the
individual tensors and then contract according to Eq.~\eqref{eq:TfGContraction},
or contract first and then fermionize the result. We must make sure that the
ordering of operations does not matter. Referring for a detailed discussion to
App.~\ref{app:bosonfermion},  we need to choose an orientation of contracted
fermionic bonds and a matching ordering of contracted bosonic indices. It turns
out that for any tensor network patch with disk topology an assignment
consistent according to the above criterion is possible. More precisely, the
order of operations is inessential up to a known  factor depending on the parity
of the number of uncontracted boundary fermion modes which does not have an
affect of bulk properties of the tensor network. We caution that  more care is
required in situations with more complex boundaries. These arise, for example,
in the calculation of $n$-point correlation functions, corresponding to $n$
additional punctures of the patch (see  Section~\ref{subsec:corr} for the
discussion of such a setting.) We finally note that in our approach the above
two-dimensional construction is key to the duality between bosonic and fermionic
systems; it here assumes a role otherwise taken by the one-dimensional
Jordan-Wigner transformations in approaches operating by dimensional reduction.

\subsection{Factorizing tensors}

As a final prerequisite for formulating our duality, we need a few added
structures: A
$\Bbb{Z_2}$\emph{-tensor} $T_\Bbb{Z_2}$  has bond dimension $2$ and is defined
by the parity condition $(T_{\Bbb{Z}_2})_{abcd}=\delta_{a+b+c+d
\,\textrm{mod}2,0}$. It is straightforward to verify that the $\Bbb{Z}_2$ tensor
satisfies the matchgate condition. In a next step, we generalize the tensor to
the presence of additional weights, $W_i$ attached to the links,
cf.~Fig.~\ref{fig:transform}(a), where the $\Bbb{Z}_2$ tensors are circles, and
the weights boxes. The latter are defined as $W_i=\text{diag}(1,w_i)$, i.e.,
diagonal matrices. This generalization, too, satisfies the matchgate condition,
with the defining matrix given by $A_{ij}=w_i w_j$. Conversely, matchgate
tensors whose Gaussian weights can be written in this way are called
\emph{factorizing}. 

While simple parameter counting shows that not every matchgate tensor can  be factorizing \footnote{For example, for tensors with four indices, we observe that a matchgate tensor has six free parameters, whereas a weighted
$\mathbb Z_2$ tensor has only four.}, they will be sufficient for our purposes. Specifically, the uniform
matchgate tensor defined by a single parameter $a$ is the simplest example of a
factorizing matchgate and its weights are $W_i=W:= \textrm{diag}(1,\sqrt{a})$.

\section{Dualities from matchgate tensor networks} \label{sec:dualities} In this
section, we will employ the bosonic and the fermionic TNs introduced above as
tools to establish  the duality between the three systems mentioned in the
introduction, the ground state of the toric code, that of the two-dimensional
class D superconductor, and the partition sum of the classical Ising model. Our
focus here, will be on the physics of the translationally invariant bulk
systems, and their respective phase transitions. The fully contracted bosonic TN
then corresponds to the partition function of the IM and likewise, the norm of a
toric code ground state with string-tension, while the fully contracted
fermionic TN evaluates to the Pfaffian of a free-fermion Hamiltonian in symmetry
class D. However, the identifications between these three systems can be made
\emph{locally} on patches of the respective tensor networks.  In
Section~\ref{sec:inhomogeneous_structures}, we take advantage of this fact and
extend the mappings to correlation functions and defect structures, starting
what could be called a dictionary between the different models dual to each
other.

\subsection*{Previous work}

To put our discussion into a larger context, we begin this section with a review
of previous studies of specific links of the duality web.

\paragraph*{IM $\to$ SC.} Previous mappings of the classical Ising partition sum
onto the SC ground state  can roughly be divided into two categories. The first
starts from a representation of the Ising partition function on a $L\times L$
square lattice as a product of $L$ transfer matrices, where each $2^L \times
2^L$-matrix represents one column of the Ising model. This formulation  stands
in the tradition of Onsager's solution \cite{Onsager}, which was subsequently
simplified by Kaufman \cite{Kaufman}, and later by Schultz, Mattis and Lieb
\cite{SCHULTZ}. One dimensional Jordan-Wigner transformations are then performed
on the  $2^L$-dimensional representation spaces of the transfer matrix to arrive
at an interpretation in terms of (Majorana) fermion ground
states~\cite{GogolinBook2004}.

A more isotropic approach has been developed by Kac and Ward \cite{KacWard} who
have expressed the partition function in terms of the determinant of a $4 L
\times 4L$ matrix by purely combinatorial considerations. This was followed by
Hurst and Green \cite{HurstGreen} who suggested a formulation using the Pfaffian
method (later refined by Blackmann and others
\cite{Blackman,PhysRevB.44.4374,PhysRevE.58.1502,PlechkoEnglish}). They noted,
that the matrix for which one computes the Pfaffian is essentially that of a
tight-binding Hamiltonian and as such can be interpreted as a free fermion
system in 2D. This approach was recast in the language of Grassmann variables by
Berezin \cite{Berezin_1969} and a formulation close in spirit to the one discussed below has been presented by Dotsenko and Dotsenko in
Ref.~\cite{Dotsenko83}. 

Finally, in more recent works \cite{Cho1997,Read_2000,Gruzberg,ChalkerMerz}, the connection between the IM and non-interacting fermions was discussed in the context of network models and a connection between network models and Gaussian fermionic tensor networks was noted in Ref.~\cite{Ludwig22}.

\paragraph*{TC $\to$ IM.} To the best of our knowledge, the duality of the IM partition sum and a TC with string tension was first explored by
Castelnovo and Chamon \cite{Castelnovo} and later generalized  in Ref.~\cite{Zhu}.

\paragraph*{SC $\to$ TC.} We are not aware of previous mappings of the TC ground
state to that of the SC, although they are of course implied by the sequence TC
$\to$ IM $\to $ SC.

\subsection*{This work}

\begin{figure}
	\centering
	\includegraphics[width=0.45\textwidth]{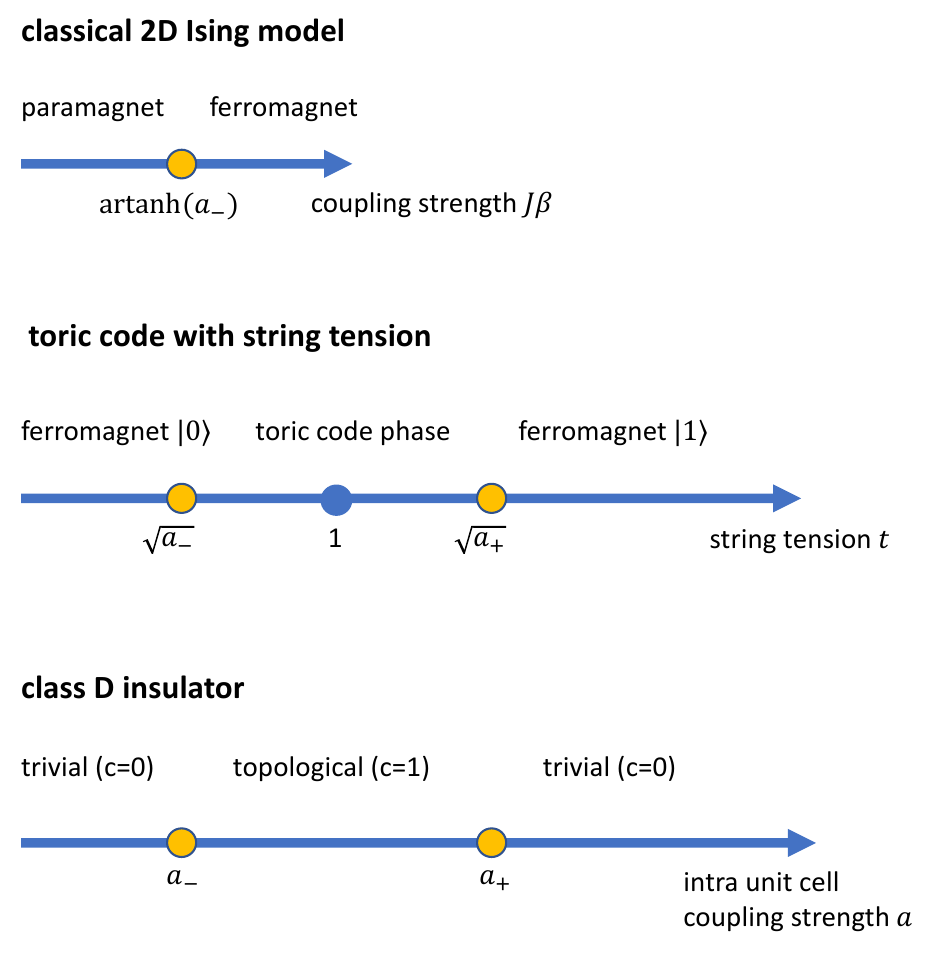}
	\caption{Phase transitions in a homogeneous single parameter matchgate tensor network interpreted in terms of three different physical systems. The control parameters $a_\pm=\sqrt{2}\pm 1$. In the lower panel, $c$ denotes the topological (Chern) index.}
	\label{fig:transitions}
\end{figure}

Our starting point in this work is a translationally invariant matchgate tensor
network on a square lattice as shown in Fig.~\ref{fig:tn}. Its tensors are
characterized by a single real parameter $a$ via $a_{ij}=a$ (see
Eq.~\eqref{eq:A}).  
All three target systems, IM, TC, SC, too, are controlled by single parameters,
to be interpreted as dimensionless coupling strength, the string tension and the
band-inversion parameter, respectively. Our discussion will show how these are
to be related to the TN parameter $a$. In each case, the parameter $a$ drives a
phase transition --- the Ising ferro-/paramagnetic transition, the transition
between a topological spin liquid and an ordered state, and the transition
between a topological and a trivial superconductor
(cf.~Fig.~\ref{fig:transitions}). While the latter two are between states of
different topological order, the first is a conventional symmetry breaking
transition.  The equivalence between these drastically different types of phase
transitions is not a contradiction since the duality transform establishing it
is intrinsically non-local.

\subsection{Classical two-dimensional Ising model}
\label{ssec:IsingDuality}

\begin{figure*}
	\centering
	\includegraphics[width=\textwidth]{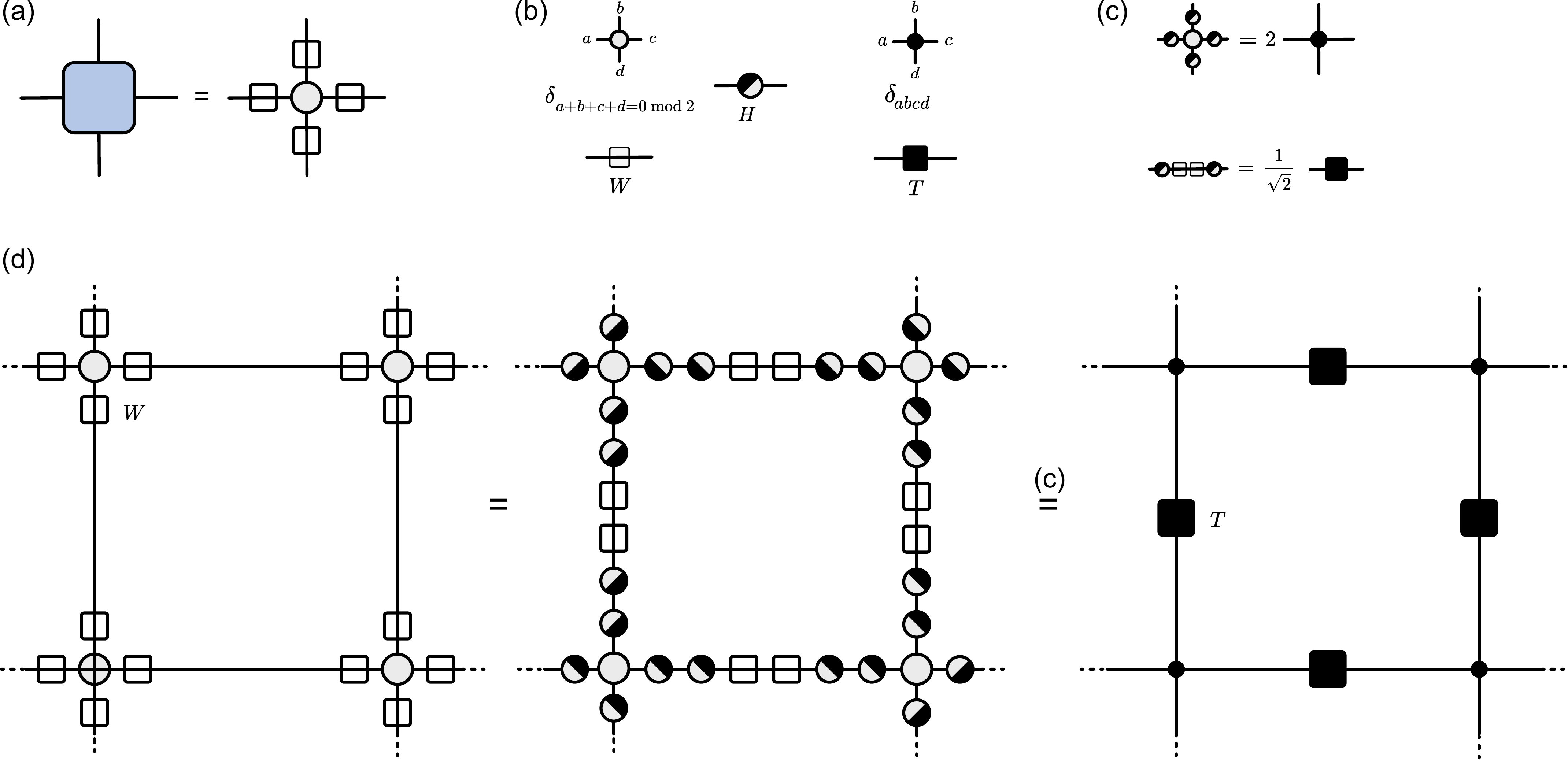}
	\caption{(a) A factorizing matchgate tensor decomposes into a $\mathbb Z_2$ tensor and weight matrices. (b) Graphical notation for the $\mathbb Z_2$ tensor (white circle), the weight matrices $W$(unfilled square), the Hadamard matrix $H$ (half filled circle, the Kronecker $\delta$-tensor (black circle) and the IM transfer matrix (black square). (c) Tensor identities used to transform a MGTN to the partition sum of the IM. (d) Transforming a weighted $\mathbb Z_2$-tensor network (left) to the Ising partition function tensor network (right) via a gauge transformation.}
	\label{fig:transform}
\end{figure*}

We start by discussing the interpretation of a factorizing  
matchgate tensor network as the partition function of the classical
two-dimensional Ising model. This connection follows from the option to realize
the even parity constraint obeyed by matchgate tensors in terms of
superpositions of weighted closed loops reminiscent of the high-temperature
expansion of the Ising partition function. Here, we derive the  equivalence
exploiting the invariance of a tensor network under basis changes in the virtual
space.

 We start from a representation of our tensors in terms of $\Bbb{Z}_2$-tensors
	with weights $W$ (see left panel in Fig.~\ref{fig:transform}(d)). Next, we
	insert products of Hadamard  matrices 
\begin{equation}
	H= \frac{1}{\sqrt{2}} \left(\begin{array}{cc} 1&1\cr 1 &-1
	\end{array} \right) 
 \end{equation}
 at all legs as shown in the center of Fig.~\ref{fig:transform}(d). With
	$H=H^\dagger=H^{-1}$ this is conceptually a gauge transformation. It is
	straightforward to verify using the relations shown in
	Fig.~\ref{fig:transform}(c) that this operation induces a transformation
	($\Bbb{Z}_2$-tensors) $\mapsto$ ($2\times \delta$-tensors)  where the
	$\delta$-tensor is defined by the condition
	\begin{equation}
 \delta_{abcd}=\delta_{ab}\delta_{bc}\delta_{cd}
  \end{equation}
  (see Fig.~\ref{fig:transform}(b)). The transformed weight matrices assume the
	form \footnote{We multiply each link by $1=\sqrt{2} / \sqrt{2}$, where the
	$\sqrt{2}$ in the numerator multiplies our $T$ and the ones in the
	denominator remove the factor of $2$ multiplying the central
	$\delta$-tensors.}
  \begin{equation} \label{eq:T}
 T:= \sqrt{2} H W^2 H=
	\frac{1}{\sqrt{2}}\left(\begin{array}{cc}
    1+a&1-a\cr 1-a & 1+a
	\end{array}     \right) .
  \end{equation}
  Assuming $a<1$, we define  
  $\beta>0$ by \eq{ J \beta = \artanh(a) }{Jofa} and $h_{00}=h_{11}=-J$ and
$h_{01}=h_{10}=J$, in order to identify $T$ with the transfer matrix of the
Ising model, 
\begin{equation}
	T=\begin{pmatrix}
		e^{-\beta h_{00}} & e^{-\beta h_{01}} \\
		e^{-\beta h_{10}} & e^{-\beta h_{11}}\end{pmatrix}.
\end{equation}
The
identification of the two non-vanishing configurations  (all links 1 or all links 0)
admitted by the central $\delta$-tensor identified with an Ising spin then implies an
equivalence of the tensor network with the classical partition sum of the Ising
model.

The two-dimensional Ising model has its magnetic phase transition at $2 J_c
\beta_c=\ln (1+\sqrt 2)$, or $a_- :=\sqrt{2}-1$, consistent with the above
assumption $a<1$ (see the upper panel of Fig.~\ref{fig:transitions}). In the
opposite case, $a>1$,  we apply a global rescaling of each bond by $a^{-1}$ to
send the  weight matrices to $W=\text{diag}(a^{-1/2},1)$. We then use the
invariance of the $\mathbb Z_2$-tensors  under a simultaneous spin flip
$\sigma_x^{\otimes 4}$ to transform to weights $W'=\sigma_x W
\sigma_x=\text{diag}(1,a^{-1/2})$, i.e., we effectively map $a \mapsto a^{-1}$.
In this case, the phase transition is at $a_+=a_-^{-1}=\sqrt{2}+1$. For $a<0$, a
global rescaling $a\mapsto -a$ shows the equivalence to the  $a>0$ parameter
domain.    
We conclude that our one-parameter family of tensor networks supports the four
parameter intervals, $a<-1, -1\le a<0, 0\le a<1$ and $1\le a$ which are
individually equivalent to the Ising models with critical points at $a=\pm
a_\pm$, respectively.

\subsection{Toric code with string tension}\label{subsec:tc} We now turn to our
second interpretation of a factorizing matchgate tensor network and show how it
is related to the wavefuntion of the toric code with string tension. (For a
direct link TC $\leftrightarrow$ IM, not using a TN intermediate, we refer to
Refs.~\cite{Castelnovo,Isakov}.) 

The \emph{toric code Hamiltonian} (without string tension)
\cite{Kitaev-AnnPhys-2003} is given by 
\begin{equation}
	H_\text{TC}= - \sum_{v \in \text{vertices}} \;\prod_{i \in v} \sigma_z^{(i)}
-\sum_{p \in \text{plaquette}} \; \prod_{i \in p} \sigma_x^{(i)},
\end{equation}
and its ground state vector $\ket{\Psi_0}$ is an equal weight superposition of
closed loops of $\ket{1}$-vectors in a background of $\ket{0}$-vectors on the
underlying square lattice. The toric code
is a most paradigmatic example of 
a $\mathbb Z_2$ spin liquid and at the same time the most studied code for topological \emph{quantum error correction} \cite{TopologicalQuantumMemory}.

\begin{figure*}
	\includegraphics[width=0.95\textwidth]{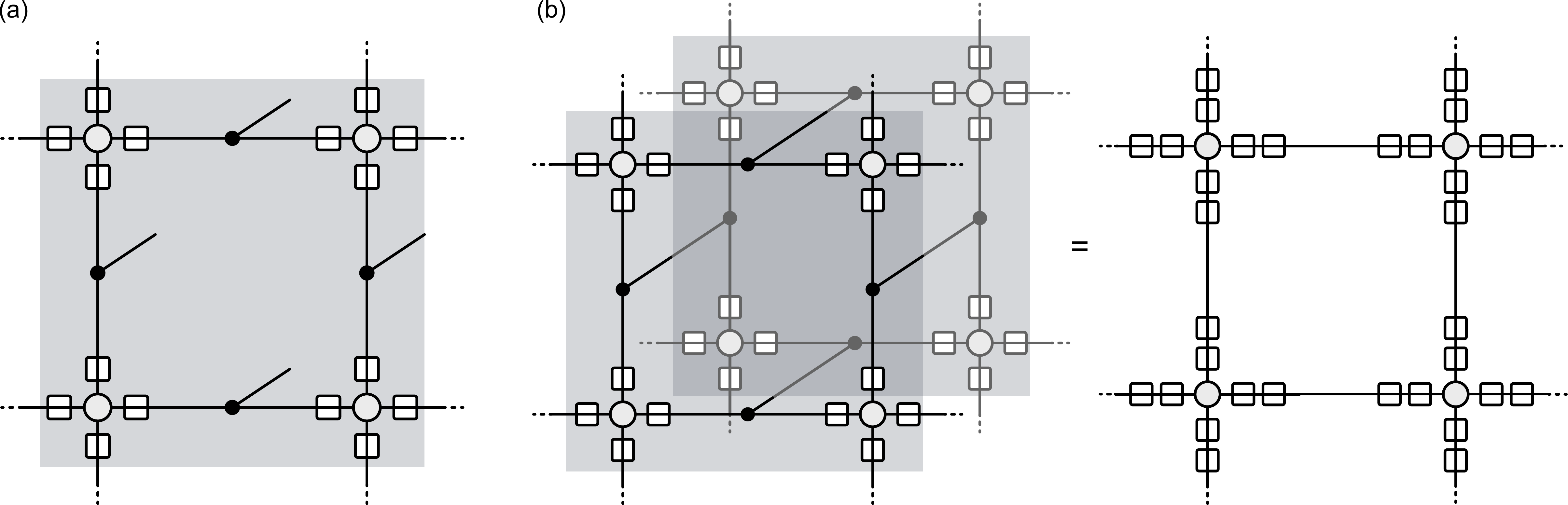}
	\caption{Toric code with string tension. (a) Ground state vector $\ket{\Psi(t)}$
	represented as a PEPS, physical indices coming out of the plane. (b) The
	overlap	$\braket{\Psi(t)|\Psi(t)}$ given by contracting the physical indices
	of the PEPS and its conjugate, represented by the mirror image.} 
	\label{fig:tc}
\end{figure*}

This state affords a  representation in terms of a simple $\mathbb Z_2$
factorizing matchgate tensor network
\cite{Verstraete_2006,Gu_2009,PEPSTopology}: First consider a configuration with
uniform $a=1$. This is equivalent to a network of $\Bbb{Z}_2$-tensors with
trivial weights. The job of the former is to admit configurations with $0,2$ or
$4$ $\ket{1}$ state vectors at each vertex, with equal weight. Summation over
all of these is equivalent to a uniform weight closed loop superposition. To
turn this sum into a quantum state, we add $3$-leg $\delta$-tensors at each
vertex (see Fig.~\ref{fig:tc}(a)). The tensor product of uncompensated physical
indices then defines the quantum ground state.     

In the toric code context, the loop sum may be turned into a weighted one by
adding string tension which penalizes or favors loops of increasing length. To
mimic this effect, we generalize the tensor network to the presence of weights
$W=\text{diag}(1,\sqrt{t})$ per half-edge defining the state vector
$\ket{\Psi(t)}$. Each  $\ket{1}$ link now comes with a factor $t$, compared to
$1$ for $\ket{0}$ links. In the extreme case of $t=0$, we obtain a trivial
ferromagnet polarized in the $\ket{0}$-state, in the opposite limit $t\to
\infty$ the system is polarized in the $\ket{1}$-state. 

The exact parent Hamiltonian of this state is given by
$H=H_\text{TC}+H_\text{ST}$ with 
\begin{equation} 
	H_\text{ST}= \sum_{v \in \text{vertices}} \prod_{i \in v} e^{- \ln t \, \sigma_z^{(i)}} \;.
\end{equation}
In the vicinity of $t \simeq 1$, the string tension term reduces to a
conventional on-site magnetic field Hamiltonian $H_\text{ST} \simeq - 2 \ln t
\sum_i \sigma_z^{(i)}$.

The critical values of the string tensions inducing a transition between the
topological spin liquid state and ferromagnetic states with $\ket{1}$ or
$\ket{0}$ polarization $t^2=a_\pm=\sqrt{2}\pm 1$ can be derived via the mapping
to the classical Ising model.
The idea is to detect a phase transition via the emergence of power law
 correlations in correlation functions  $C_{t,\alpha \beta}=\braket{\Psi
 (t)|O_\alpha O_\beta|\Psi(t)}$, where $O_{\alpha}$ and $O_\beta$ are local
 operators of the spin variables at sites $\alpha$ and $\beta$. The operator
 insertion between two states means that we are now dealing with a two layered
 tensor network where all physical indices except for those at $\alpha,\beta$ of
 that representing $\ket{\Psi(t)}$ are contracted with that representing
 $\bra{\Psi(t)}$ (see Fig.~\ref{fig:tc}(b)). The consequences of this almost
 complete contraction are detailed in Appendix \ref{app:tc} and can be
 summarized as follows: the contraction of the $\delta$-tensors sitting on the
 bonds effectively causes a collapse of double bonds (representing bra- and
 ket-state) to a single one. On this effective bond, we have the square of two
 amplitude weights, i.e., the effective weight $W^2=\text{diag}(1,t)$. This is
 equivalent to a matchgate tensor with uniform $A$ matrices for which
 $a_{ij}=t^2$ and implies criticality at $t^2_\pm=a_\pm$, in agreement with the
 values reported in Refs. \cite{Castelnovo,Isakov,Zhu} (see middle panel of Fig.~\ref{fig:transitions}).

\subsection{Class D superconductor} \label{subsec:classD}

Our third construction establishes a connection to the topological SC. For the
convenience of readers not well-versed in the physics of topological
superconductivity, a brief review is included in Appendix
\ref{sec:ClassDReview}. The punchlines of this discussion are that (a) the free
fermion Hamiltonian of a class D superconductor affords a representation in
terms of a \emph{Majorana bi-linear form}
\begin{align}
	\label{eq:Majoranabi-linear}
	\hat{H}:= i \Theta^T H \Theta,
\end{align}
where $H$ is the first quantized Hamilton matrix, and the components of $
\Theta$ Majorana operators, to be identified as real and imaginary parts of
complex fermion creation and annihilation operators. (b) In a translationally
invariant system, the eigenstates of $H$ define Bloch bands, labeled $n$,  which
individually carry Chern numbers $c_n$. (c) Transitions between states of
different topology change these numbers (at a conserved total number $\sum_n
c_n=0$) and are signaled by a change of the $\Bbb{Z}$-valued  topological index $c=\sum_{n=1}^N c_n$ over Chern numbers carried by
individual Bloch bands, $n=1,\dots,N$, with band energies below the
superconductor band gap at $\epsilon=0$.
(d) Such changes of integer invariants require  touching of the bands $
n$ and $n+1$ involved in the change of Chern numbers. In the vicinity of these
hotspots in the Brillouin zone,  the local Hamiltonian may be approximated by a
\emph{two-dimensional Dirac Hamiltonian}
\begin{align}
	\label{eq:DiracHamiltonian}
	H^{(2)}= \kappa (q_1 \sigma_1 + q_2
\sigma_3 + (m+\alpha q^2) \sigma_2)+\mathcal{O}(q^3),
\end{align}
where $\kappa$ is an overall real constant, $q=(q_1,q_2)^T$ is the momentum
difference from the band touching point, $m$ a mass parameter measuring the
distance to the critical point, and $\alpha$ a parameter entering the second order
expansion in the local dispersion relation. In this representation, the
assignment of Chern numbers reads $(c_n,c_{n+1})=(0,0)$ for $m\alpha>0$, and
$(+1,-1)$ for $m\alpha<0$, i.e., the topological index is determined by the sign of
the mass gap. (Exchange $q_1 \leftrightarrow q_2$ corresponds to the sign change
$(-1,+1)$.)

\begin{figure}[t]
	\centering
	\includegraphics[width=0.3\textwidth]{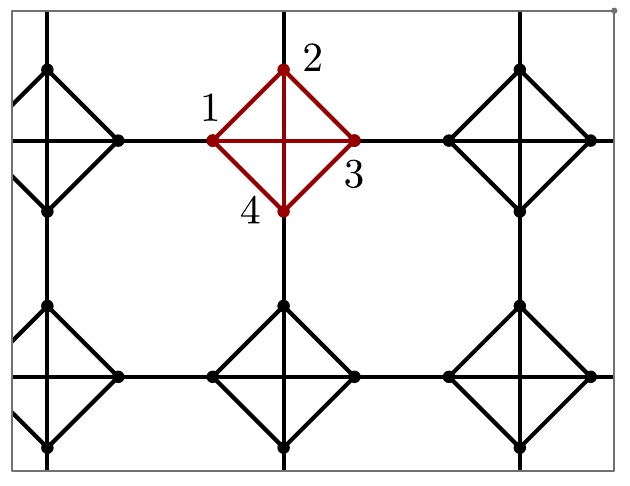}
	\caption{Lattice structure comprising four-site unit cells with all to all
	connection of uniform strength $a$ extended to a square
	lattice.\label{fig:TNLattice}}
\end{figure}

\begin{figure}[h]
	\centering
	\includegraphics[width=0.45\textwidth]{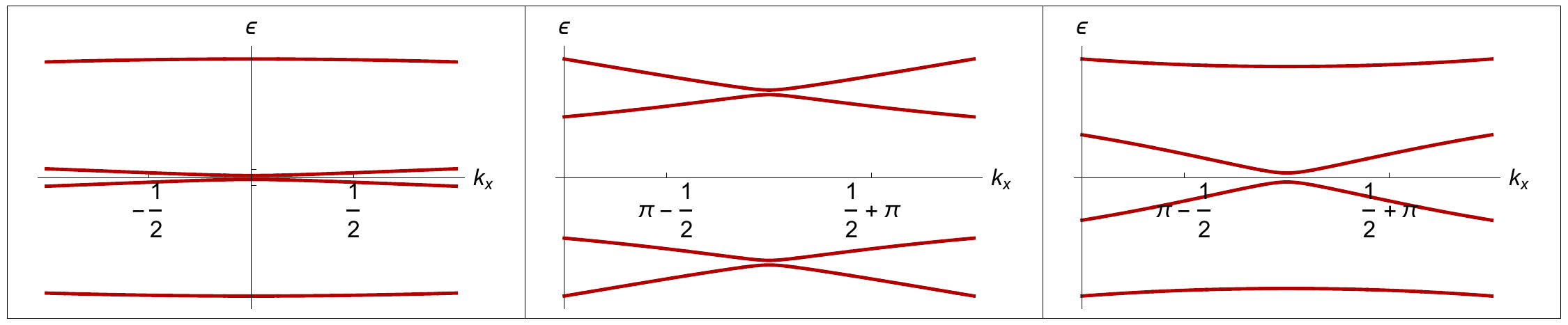}
	\caption{A cut at $k_y=0,\pi,\pi$ (left, center, right) through the
	two-dimensional dispersion relation at parameter values $a=a_++0.3, b+0.04,
	a_-+0.02$ close to the topological phase transition points. Notice how the
	transition at $b$ involves the formation of a Dirac point between the
	occupied bands $c=1,2$.  \label{fig:Dispersion}}
\end{figure}

We now establish a connection to the fermionic representation of our TN by
identifying the bilinear form Eq.~\eqref{eq:Majoranabi-linear} with
Eq.~\eqref{eq:HCA}, with the matrix defining a matchgate tensor network on a
square lattice with  tensors of uniform weight, $A^\alpha_{ij}=a$. Of course,
this connection remains formal unless contact with the momentum space topology
of a superconductor is made. To this end, note that the TN structure includes
unit cells comprising four Majorana fermions with all to all connection of equal
strength $a$. Connecting these cells to a square network (see
Fig.~\ref{fig:TNLattice}), and switching to a momentum space representation,
$\tilde H = \sum_k \Theta_k H_k \Theta_{-k}$  the system is represented in terms
of the effective four band Hamiltonian
\small
\begin{align}\nonumber 
 	H=&i \, a\left( \begin{matrix}
		0&1&1&1\cr 
		-1&0&1&1\cr 
		-1&-1&0&1\cr 
		-1&-1&-1&0 
	   \end{matrix} \right)+i \left( \begin{matrix}
		0&0&e^{ik_x}&0\cr 
		0&0&0&e^{ik_y}\cr 
		-e^{-ik_x}&0&0&0\cr 
		0&-e^{-ik_y}&0&0 
	   \end{matrix} \right).
\end{align}
\normalsize 
The numerical brute force computation of Berry curvatures for this Hamiltonian
reveals a sequence of six topological phase transitions at $a$-values
$(\sqrt{2}-1,1, \sqrt{2}+1))=:(a_-,1,a_+)$ and the negative of these. Labeling
the energy bands $1,2,3,4$ in ascending order in energy, we have $N=2$ occupied
bands $n=1,2$ with a pattern of Chern numbers shown in Table
\ref{tab:ChernNumbers}: Starting from a topologically trivial phase at $a>a_+$,
a phase transition involving bands $n=2,3$ at $a=a_+$ (see
Fig.~\ref{fig:Dispersion}) defines the entry into a topological superconductor
phase with $c=1$. At $a=1$, a phase transition involving the topological order
of bands $1,2$ (and $3,4$) leads to a redistribution of band Chern numbers
without changing the total topological index, $c$. Finally, at $a=a_-$, we have
a transition back to $c=0$, however, this vanishing value is  nontrivial in that
it implies two bands with mutually canceling non-vanishing index, $c_1=1$ and
$c_2=-1$.

To obtain an explicit low energy reduction of the system near, say, the critical
point $a=a_-$, we verify that at the momentum hot-spot $k=(\pi,\pi)$  our
Hamiltonian has two zero eigenvalue states, 
\begin{eqnarray}
    v_1 &:=&\frac{1}{2}(-\sqrt{2},-1,0,1),\\
    v_2 &:=&\frac{1}{2}(0, 1,\sqrt{2},1).
\end{eqnarray}
Setting $a=a_-+{m}/({2(1+\sqrt{2})})$, and $k_i=\pi+q_i$, the two-band
reduction, $H^{(2)}$, of the Hamiltonian obtained by projection onto the space
spanned by $v_{1,2}$ assumes the form of the Haldane
\cite{PhysRevLett.61.2015} 
Chern insulator  
\begin{align}
	H^{(2)}=-\frac{1}{2}&\left( \sin(q_2)\sigma_3+\sin(q_1)\sigma_1 \right. \\	
	&+\left. (2+m-\cos(q_1)-\cos(q_2))\sigma_2 \right).
 \nonumber
\end{align}     
For $m>0$ and $m<0$,   the two bands associated to this Hamiltonian carry
winding (Chern) numbers $(0,0)$ and $(-1,1)$, respectively. The low energy Dirac
approximation Eq.~\eqref{eq:DiracHamiltonian} (with $\kappa=-\frac{1}{2}$) is
obtained by expansion in $q$ up to second order.  

\begin{table}[h]
	\centering
	\begin{tabular}{c| c c c c}
		 
		 $a \in $&  $(0,a_-)$ & $(a_-,1) $& $  (1,a_+)$& $(a_+,\infty)$\cr
		 \hline \hline
		$c_4$ & -1 & -1 & 0 &0 \cr $c_3 $ & 1 & 0 & -1 &0 \cr $c_2$ & -1 & 0 & 1
		&0 \cr $c_1 $ & 1 & 1 & 0 &0 \cr \hline
		$c $ & 0 & 1 & 1 &0 \cr 
	\end{tabular}
	\caption{Chern numbers carried by the four bands of the system for positive
	values of the parameter $a$. (The pattern is symmetric around $a=0$ and in
	this way can be extended to negative values.) The phase transitions changing
	the topological index, $c$, occur at $a_\pm$. \label{tab:ChernNumbers}}
\end{table}

With these identifications, the equivalence between the TN and the SC is
established, and from here one may --- via the non-local boson-fermion mapping
outlined in Section~\ref{sec:MG} --- pass to the bosonic systems TC and IM.
Notice how in our discussion of the fermion system, the emphasis shifted from
real space to momentum space structures. Nevertheless, the full microscopic
structure of the system remains under control, and this will become essential in
the next section when we turn to the discussion of correlation functions.

\section{Extension to inhomogeneous
structures}\label{sec:inhomogeneous_structures}

So far, we have looked at the duality between our three systems in the
translationally invariant case. However, they all show rich behavior when
translational invariance is broken by domain walls or other defect structures,
examples including  vortices binding Majorana zero modes in the
SC~\cite{FuKane2008}, non-local `disorder' operators describing the correlations
between endpoints of defect links in the IM~\cite{KadanoffCeva}, anyonic
excitations as endpoints of error strings in the TC~\cite{Kitaev-AnnPhys-2003},
or the long-rangedness of correlation functions at criticality (where we
consider a correlation function as the result of an insertion of infinitesimally
weak probing inhomogeneities.) The general duality must include a mapping
between these defect structures, and their accompanying correlations. However,
we anticipate that the passage from fermionic to bosonic systems involved in
going from the SC to the IM or the TC will introduce  an element of
non-locality: the correlation between objects that are local in one setting may
turn into the insertion of non-local string like objects in another. 

These complications find their perhaps most vivid manifestation in the duality
mapping of the simplest correlation function probing the superconductor, that
between two Majoranas, Eq.~\eqref{eq:MajoranaTwoPoint} below. The dual
representation of this function in the Ising context is famously complex
\cite{KadanoffCeva,Dotsenko83} and it involves the pairing of a spin and a
disorder operator to a hybrid operator. The simultaneous appearance of a local
(spin) and a non-local (disorder) operator reflects that we are representing the
correlations of a fermion in the language of a bosonic model. Specifically, the
spin-disorder dual of the Majorana correlation function responds sensitively to
the relative placement of the two compound operators, as discussed in
Ref.~\cite{KadanoffCeva} and explored explicitly in a dual fermionic description
in Ref.~\cite{Dotsenko83}.

In the following, we show how the tensor network allows one to map defect
structures and correlation functions with maximal explicitness. We will
illustrate the construction on two examples. The first is the mapping of the
above SC Majorana two-point correlation function. We will construct local pairs
of spin and disorder operators as fractionalized representatives of the
Majorana, and address the importance of their relative ordering on the lattice.
The second example is motivated by the question what form these structures
assume in the TC language. We find that for general parameter values the answer
assumes the form of an exponentially decaying and not very illuminating ground
state operator expectation value. However, the analysis of the latter becomes
more rewarding once we introduce a spatial domain wall, i.e. an object which in
superconductor language defines the surface of a topological insulator. In this
case, our correlation function describes the spatial extension of a topological
boundary mode, and we will discuss how this object turns into a non-decaying
string operator expectation value in the toric code.

\subsection{Correlation functions} \label{subsec:corr}

\begin{figure}
	\centering
	\includegraphics[width=0.45\textwidth]{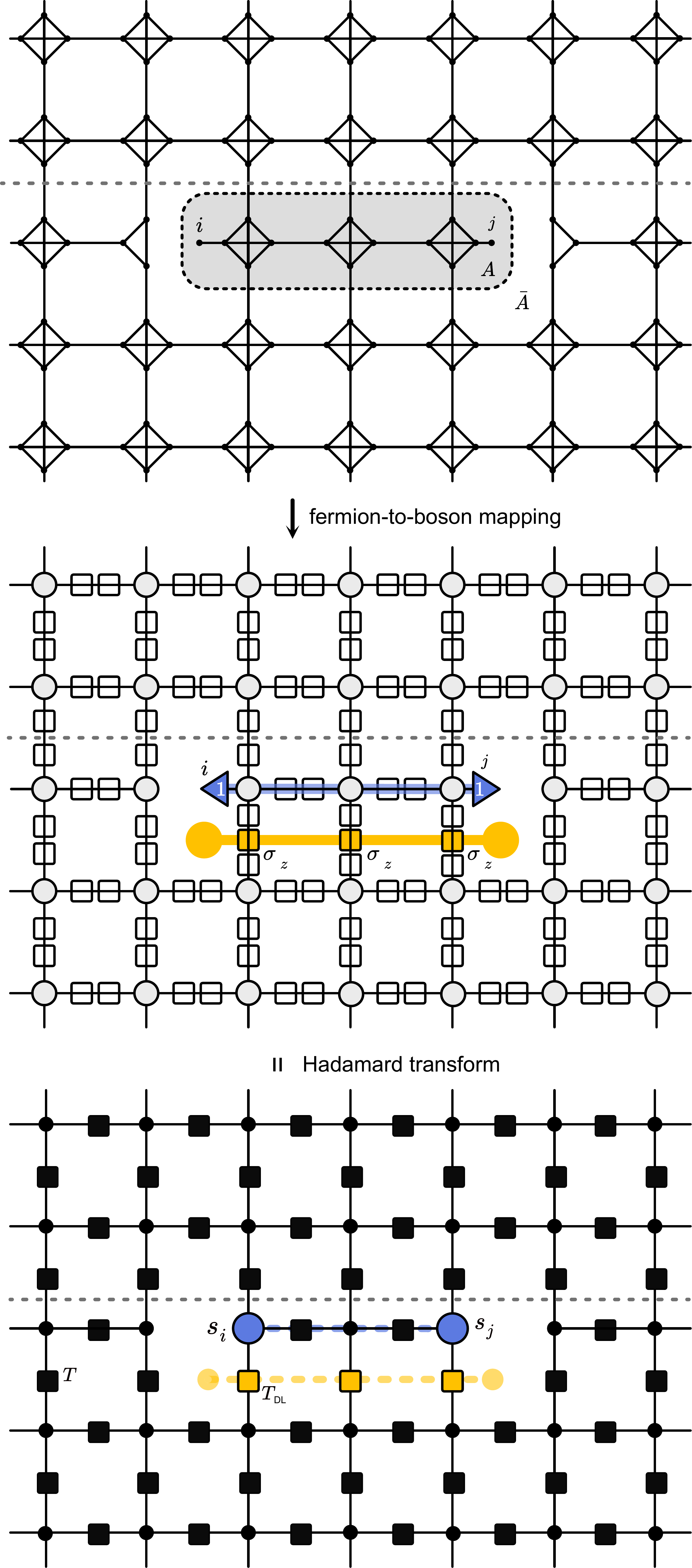} 
	\caption{Top. Fermionic TN for $\langle \theta_i \theta_j \rangle$ divided
	into fermion parity even regions $A$ and $\bar A$. Center. The TN after
	fermion-to-boson mapping. We obtain a bosonic weighted $\mathbb Z_2$ TN with
	$\sigma_z$ matrices (transparent yellow squares) acting along a defect line
	on the dual lattice (yellow) and projections onto the state vector $\ket{1}$
	at the sites $i$ and $j$, respectively. Bottom. The TN of the IM partition
	function obtained via a Hadamard gauge transformation. The
	$\sigma_z$-matrices are transformed to transfer matrices $T_\text{DL}$ of
	inverted coupling strengths $J\to -J$ (filled yellow squares) along the
	defect line and the spins at site $i$ and $j$ (blue dots) contribute a sign
	factor to the partition function leading to the expression in
	Eq.~\eqref{eq:isingcorr}. The dashed gray line represents a (potential)
	boundary between regions of different $a$-values discussed in
	Sec.~\ref{subsec:boundary}. }
	\label{fig:corr_order}
\end{figure}

Correlation functions are described by a TN modified at the two points between
which correlations are measured. These modifications can interfere with the
fermionization procedure discussed in the previous section. Specifically, if the
observables under consideration are of odd fermion parity, the mapping between
bosonic and fermionic representations introduces a  string-like object
connecting the two point observables. To illustrate the principle -- and to be
entirely concrete -- we start out from the fermionic two point correlation
function 
\begin{equation}
	\label{eq:MajoranaTwoPoint}
	\langle \theta_i \theta_j \rangle := \int (\dd \theta)_C \, \theta_i \theta_j \,e^{\frac 1 2 \theta^T (A+C) \theta} \;,
\end{equation}
where $C$ denotes the signed adjacency matrix of the network and $A$ the
collection of all matchgate tensor generating matrices. We may think of this as
the contraction of a fermionic tensor network subject to two incisions as
illustrated in Fig.~\ref{fig:corr_order}. Doing the integral, we obtain
\begin{equation}
	\langle \theta_i \theta_j \rangle \propto H^{-1}|_{ij}\;, \label{eq:hinv}
\end{equation}
with the Hamiltonian $H=\mathrm i (A+C)$. Close to a critical point, e.g.,
$a=a_-$, $H$ can be approximated by the  two-band Hamiltonian
Eq.~\eqref{eq:DiracHamiltonian} linearized around $q_x,q_y=0$. At criticality,
$m=0$, the gap vanishes and the correlation function approaches an $l^{-1}$
power law where the exponent follows from dimensional analysis and  $l$ is the
distance between  $i$ and $j$. However, in addition to this asymptotic distance
behavior, we have additional short range lattice structures to consider. A
single point $i=((x,y)b)$ is defined by unit cell coordinates $(x,y)$ and an
intra-cell index $b=1,\dots, 4$. It turns out that only pairs $i,j$ with select
combinations of this data survive projection onto the two-band reduction, and
hence are long-range correlated. For example, considering  points  separated
along the $x$-direction, we find $\langle
\theta_{(x,y)1}\theta_{(x+l,y)1}\rangle =0$, while $\langle
\theta_{(x,y)1}\theta_{(x+l,y)3}\rangle \propto 1/l$. 

\paragraph*{Fermion boson mapping.} We next aim to represent this correlation
function in the language of the bosonic TN. The challenge here is the presence of
fermion parity odd tensors at sites $i$ and $j$.  To deal with this situation,
we decompose the tensor network into two parts which are individually
fermion-parity even. The first of these,  $A$, contains $\theta_i$ and
$\theta_j$, and hence is fermion even in total. The complement, $\bar A$,
contains the rest of the tensor network. For the sake of simplicity, we chose
$A$ to be as small as possible, namely as a chain of tensors connecting sites
$i$ and $j$ (see Fig.~\ref{fig:corr_order} top). Referring to Appendix
\ref{app:bosonfermion2} for a more detailed discussion, the goal now is to
contract all tensors in $A$ and reorder the fermionic modes into standard
ordering. (The tensors of $\bar{A}$ can be assumed to be in standard ordering to
begin with.) As a necessary byproduct, this operation introduces a string of
fermion parity tensors between sites $i$ and $j$. In a final step, we  
contract the bosonic versions of $A$ and $\bar A$ to obtain a bosonic tensor
network in standard ordering.  

In bosonic language, the above  parity string is expressed through
$\sigma_z$-matrices acting on the virtual bonds of the TN (see
Fig.~\ref{fig:corr_order} center). The  $\theta$-modes themselves become
projections onto  spin-up, i.e., $\ket{1}$-state at sites $i$ and $j$. The
resulting bosonic tensor network is shown in the middle panel of
Fig.~\ref{fig:corr_order}. Having defined two alternative representations of the
TN subject to point sources at sites $i$ and $j$, we now turn to the
interpretation of these structures in terms of condensed matter correlation
functions.

\textbf{SC:} Eq.~\eqref{eq:MajoranaTwoPoint}  affords an obvious interpretation
as the Majorana ground state correlation function of a superconductor.
Specifically, we think about the r.h.s. of Eq.~\eqref{eq:hinv} as the ground
state, or  zero energy, $\epsilon=0$,  matrix element $G_{ij}$ of the resolvent
$G=(\epsilon - H)^{-1}$. (Due to the presence of a spectral gap and the absence
of convergence issues in the Majorana functional integral, it  is not necessary
to shift $\epsilon$ into the complex plane as for generic Green functions.) Its
algebraic decay then reflects the long-range correlation of Majorana
quasiparticles at the gap closing transition of the topological superconductor.

\textbf{IM:} Next, we interpret the  bosonic incarnation of the sourced tensor
network in terms of an Ising model correlation function. As before,  we obtain
the corresponding partition sum  by performing a Hadamard transform (see
Fig.~\ref{fig:transform}) which in the presence of sources has the following
effect: the $\ket{1}$-projections become projection to the $\ket{-}$-state,
meaning, if the spin at position $i$ is up, we obtain a minus sign. The same
holds for the spin at position $j$. We identify the parity string as a
\emph{defect line} (DL) and observe that the transfer matrices at all bonds
crossed by that defect line are given by
\begin{equation}
	T_\text{DL}=\sqrt{2} H W^2 \sigma_z H = \frac{1}{\sqrt{2}} \begin{pmatrix}
		1-a & 1+a \\ 1-a &1+a
	\end{pmatrix} \;.
\end{equation}
Comparing this to the result in Eq.~\eqref{eq:T} and Eq.~\eqref{eq:Jofa}, we
note that these are transfer matrices with inverted coupling strength, i.e.,
$T_\text{DL}=T_{J \to -J}$.

In summary, the tensor network after Hadamard transform shown at the bottom of
Fig.~\ref{fig:corr_order} is a spin-spin correlation function of sites $i$ and
$j$, but with a Hamiltonian $H_\text{DL}$ that has a modified coupling strength
$J\mapsto -J$ along a defect line
\begin{equation}
Z_{\text{corr},ij}= \sum_{\{s_i\}=\pm 1 } s_i s_j e^{-\beta H_\text{DL}(\{s_i\})}\;.\label{eq:isingcorr}
\end{equation}
The specific path entering the construction  of the defect line depends on the
arbitrary choice made in dividing the tensor network into $A$ and $\bar A$.  

These building blocks entering the construction of the Ising representation of
the fermion correlation function have a long history of research. Specifically,
$\sigma_z$ string  line defects extending from a point of the lattice along
arbitrary paths to infinity are called \emph{disorder
operators} and and have been introduced in Ref.~\cite{KadanoffCeva}. They owe their
name to the fact that they assume finite expectation values in the disordered
high-temperature phase of the IM, and they are related by (Kramers-Wannier)
duality to the native local spin operators. Composite correlation functions
involving pairs of disorder operators (now connected by a finite defect line)
and spin operators were considered in the same reference, where the importance
of the precise relative positioning of spin and disorder operator (addressed
above in the language of the Majorana representation) was discussed. In a
lattice construction conceptually close to the present one but formulated
directly with the framework of the IM, Ref.~\cite{Dotsenko83} investigated the
braiding properties of the composite operator to demonstrate that it defines an
effective (Majorana fermion). Within our present construction the bridge between
free Majorana fermions and composite IM operators is established in the explicit
and arguably maximally concise mapping of the fermionic to the bosonic TN.  

\begin{figure}
	\centering
	\includegraphics[width=0.35\textwidth]{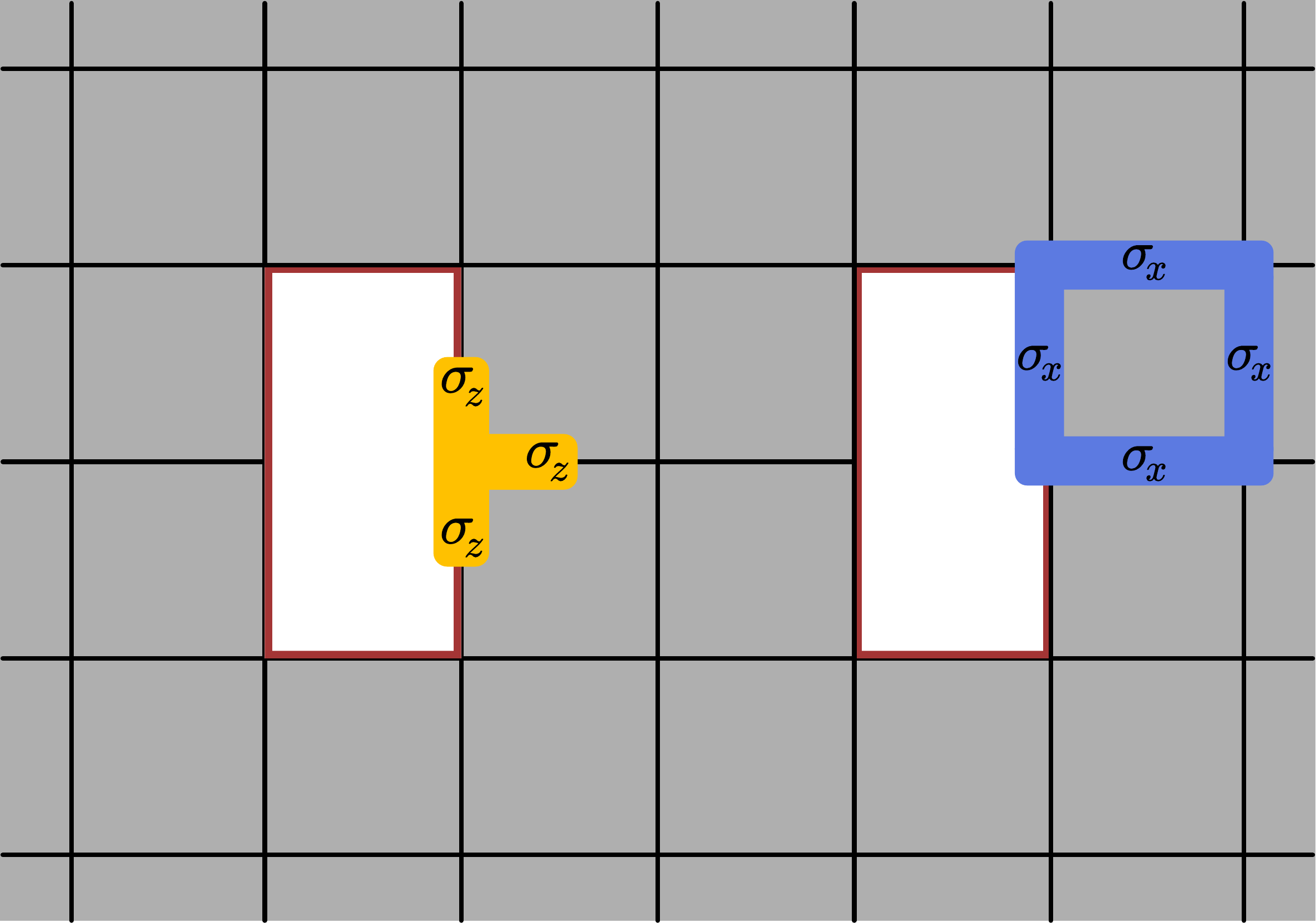}
\caption{Toric code with smooth boundaries. The Hamiltonian in the bulk (shaded in grey) is given by the conventional square lattice Hamiltonian. Along the boundary all vertex operators are modified such that they are given by a product of $\sigma_z$-matrices acting on the edges inside the grey shaded region only. The plaquette operators remain the same.}
	\label{fig:smooth}
\end{figure}

\begin{figure}
	\centering
	\includegraphics[width=0.35\textwidth]{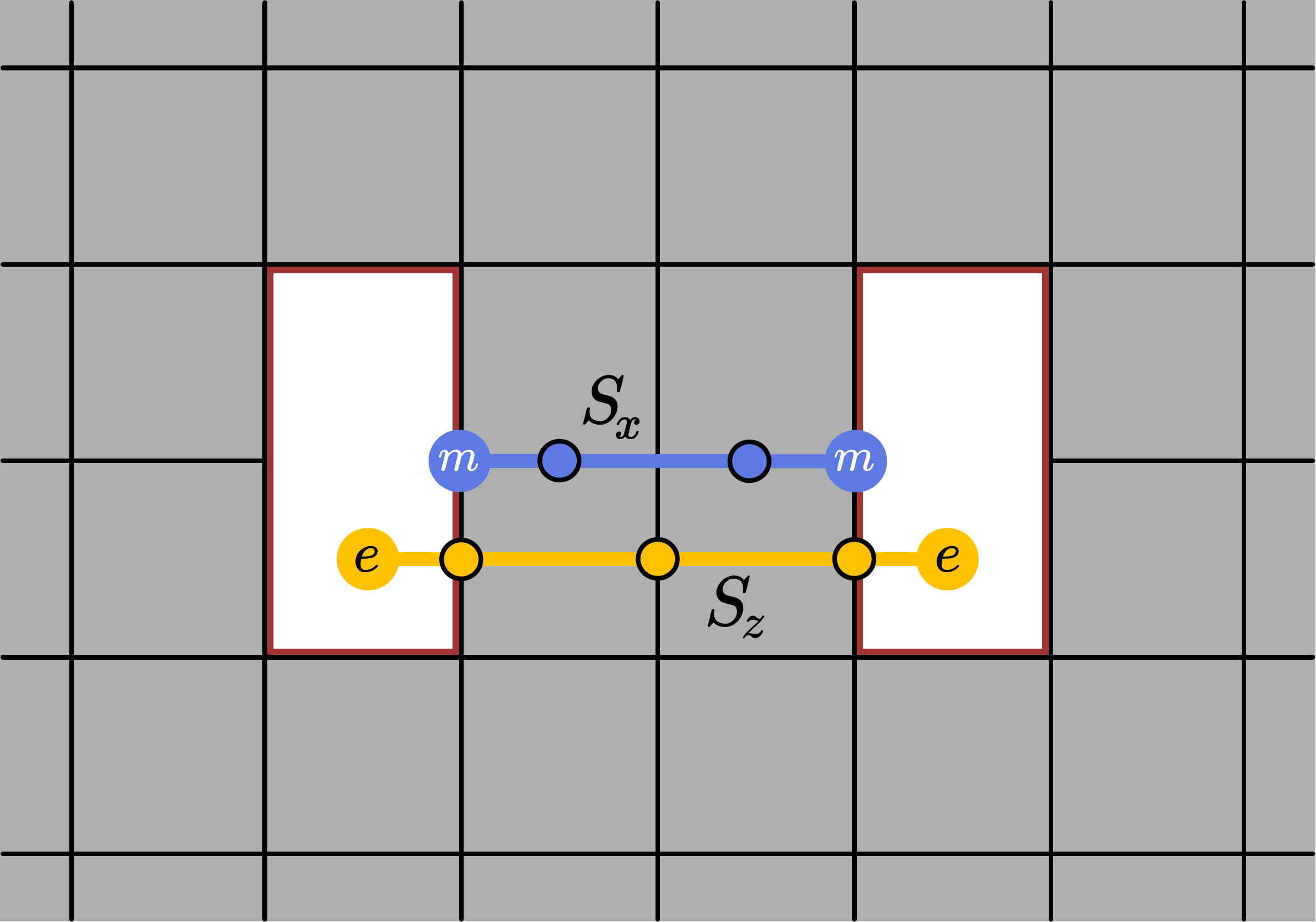}
\caption{Toric code with smooth boundaries around two  holes that arise from the
correlation function `incisions' at sites $i$ and $j$
(cf.~Fig.~\ref{fig:corr_order} center). The string of $\sigma_x$-matrices, $S_x$
(blue), creates parity violations ($m$-excitations) at its end-points. The
string of $\sigma_z$-matrices, $S_z$ (yellow), connects the incision sites and
creates $e$-excitations at its endpoints.}
	\label{fig:tc_corr}
\end{figure}

\textbf{TC:} We finally turn to the an interpretation of the correlation
function in terms of toric code ground state expectation values. To be more
precise, we consider the bosonic TN in the middle panel of
Fig.~\ref{fig:corr_order} and try to identify a toric code ground state
$\ket{\Psi}$ such that $\braket{\Psi|\ldots |\Psi}$ corresponds to the given TN
locally and the ellipses stand for appropriate operators to be identified as
well. For general $a$-values, this procedure leads to complicated and not very
revealing expressions. However, at the toric code fixed point, $a=1$, i.e. in
the absence of string tension,  the construction becomes straightforward.
Focusing on this case, we note that for $a=1$ the weight matrices are identity
matrices, meaning that we can  omit them in the pictorial representation of the
TN in the middle panel of Fig.~\ref{fig:corr_order}.

We next note that the  TN modified for the presence of source terms has a number
of specific features which help to identify the state vector $\ket{\Psi}$ and the
operators featuring in the expectation value. First, it is
missing a bond to the left (right) of site $i$ ($j$). This missing bond suggests
to choose $\ket{\Psi}$ as the ground state vector of the toric code with holes at sites
$i$ and $j$. In the context of the toric code, the minimal surface bounding these holes is known as  a \emph{smooth boundary}. The
ground state of a toric code containing smooth boundaries is a  superposition of all closed
loops of spin-up states \footnote{By contrast, the ground state of a TC with  
 \emph{rough boundaries} where bonds poke into the vacuum contains strings ending there.}, where the loops are allowed to include  boundary links
(cf.~Fig.~\ref{fig:smooth}). The remaining features of the TN we need to
reproduce are the string of sign flips along the defect line, and the
$\ket{1}$-projections at sites $i$ and $j$. 

The constructive modeling of these structures is a nice illustration of  tensor
network constructions and detailed in App.~\ref{app:tc}. We here simply state
the result and note that the state we are looking for is $\ket{\Psi}:= S_x
\ket{\Psi_\square}$, where $\ket{\Psi_\square}$ is the ground state of the TC
with smooth boundaries mentioned above and $S_x=\sigma_x \otimes \ldots \otimes
\sigma_x$ is a string of $\sigma_x$-matrices (bit flips) between sites $i$ and
$j$. Applied to the ground state, this operator generates an excited state with
two $m$-anyons located at the endpoints of the string
(cf.~Fig.~\ref{fig:tc_corr}) \cite{Kitaev-AnnPhys-2003}. The overlap
$\braket{S_x \Psi_\square | S_x \Psi_\square}$ reproduces the
$\ket{1}$-projection of the TN in the middle panel of Fig.~\ref{fig:corr_order}.

The sought after operator replacing the ellipses in $\braket{\Psi|\ldots|\Psi}$
is $S_z=\sigma_z \otimes \ldots \otimes \sigma_z$ -- a string of
$\sigma_z$-matrices (phase flips) along a path connecting sites $i$ and $j$ on
the dual lattice. This operator is known to generate an excited state with
$e$-anyons at its endpoint (cf.~Fig.~\ref{fig:tc_corr})
\cite{Kitaev-AnnPhys-2003}. Here, the $e$-anyons lie inside the two holes with
smooth boundary and instead give rise to another (orthogonal) ground state of
the Hamiltonian with smooth boundary. The expectation value of this operator for
a TC ground (or excited) state reproduces the string of sign flips in the TN.

In summary, the expectation value reducing to the TN in the middle panel of
Fig.~\ref{fig:corr_order} and satisfying the above locality condition is given
by $\braket{\Psi|S_z|\Psi}=\braket{S_x \Psi_\square |S_z |S_x \Psi_\square}$.
Since the support of $S_x$ and $S_z$ can be chosen to be non-overlapping, these
two operators commute. With $S_x^2=\id$, we find
$\braket{\Psi|S_z|\Psi}=\braket{\Psi_\square | S_z \Psi_\square}$. 

Let us discuss this result. We have seen that the presence of the $S_x$ strings
is irrelevant in the expectation values. The job of these operators was to
implement the $\ket{1}$-projections, which in turn were dual to the spin
operators in the Ising context. Since $a=1$ translates to
$J\beta = \infty$ (cf.~Eq.~\eqref{eq:Jofa}) this dual IM is deeply within the
ferromagnetic phase, where spins are uniformly aligned and hence drop out of correlation functions.
We next note  that the operator $S_z$ applied to the ground state generates a
state with two $e$-anyons at its endpoints. This state is orthogonal
to the state $\ket{\Psi_\square}$ and the overlap $\braket{\Psi_\square | S_z \Psi_\square}$ is trivially
zero. This, again, is expected by duality -- the $S_z$ operator corresponds to the disorder correlation function in the IM which vanishes in the ferromagnetic
phase. 

We thus conclude that all the so far effort has led to the  dual TC
representation of a trivially vanishing correlation function. This vanishing is
due to the fact that we are considering correlations on top of a  bulk
background which, for $a=1$, is fully gapped. However, the situation becomes
more interesting, when we allow for the presence of a system boundary spatially
aligned to our probe operators.

 \subsection{Boundary phenomena}

\label{subsec:boundary}

Turning back to the TN, assume a separation of our system into a `topological'
region, $R$ defined by a value $a>a_-$ of the coupling constant, and a
`non-topological' complement with $a<a_-$. In SC language, this will define an
interface between a topological and a non-topological superconductor, which we
know supports gapless Majorana boundary modes. On this basis, we anticipate that
the TC representation, too, will support long range correlations, which we can
probe by putting two observation points $i$ and $j$ next to it. This is the
setup considered in the following. 

 Consider the TN in the middle panel of Fig.~\ref{fig:corr_order} where now the
weight matrices above the dashed line have $a<a_-$. For simplicity, we set
$a=0$, which reduces these matrices to projections onto the state vector
$\ket{0}$. Below the dashed line, where $a>a_-$, we choose  $a=1$, implying that
the weights become unit matrices. In effect, this network has all bonds above
the dashed line, including those that are crossed by it, removed. This defines a
maximally simple interface between a topological region and `vacuum'. 

\textbf{TN:} In TN language, the above construction introduces another smooth
boundary along the interface, in addition to that surrounding the correlation
function observation points. As before, we aim to identify the correlation
function generalized for the presence of the interface with a suitable ground
state expectation value. Reconsidering the construction in the previous section,
we conclude that it remains unchanged, only that the ground state in question
now is that of the system with the generalized boundary,
$\ket{\Psi_{\square,|}}$. In the expectation value  $\braket{S_x
\Psi_{\square,|}|S_z|S_x \Psi_{\square,|}}$, the $S_x$ string is irrelevant as
before, leading to $\braket{\Psi_{\square,|}|S_z| \Psi_{\square,|}}$.

\begin{figure*}
	\centering
	\includegraphics[width=\textwidth]{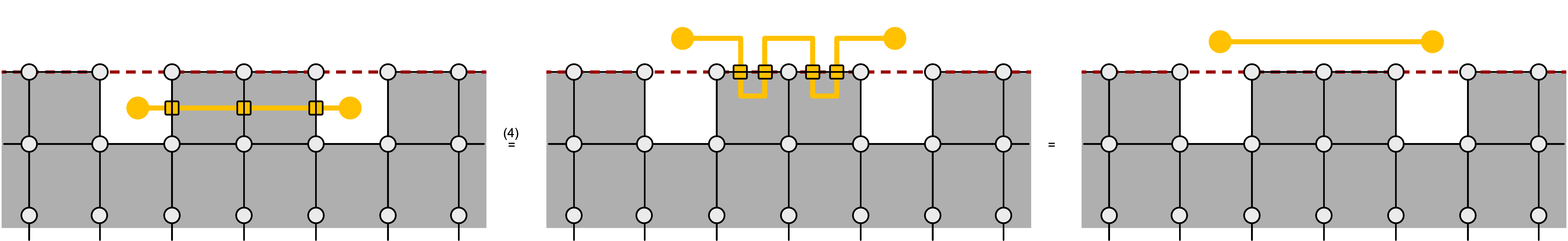}
\caption{Deforming and removing a $S_z$-string at a smooth boundary by repeated
application of identity (4) from Fig.~\ref{fig:z2_id}.}
\label{fig:zboundary}
\end{figure*}

So far, we have not specified the positioning of the sites  $i$ and $j$ relative
to the boundary. The situation gets interesting when they come close to it (as
depicted in the middle panel of Fig.~\ref{fig:corr_order}). Once they touch,
the holes surrounding $i$ and $j$ are 'cut open' and partially lie
\emph{outside} the bulk. This has a dramatic effect. While previously, we found
that $\braket{\Psi_{\square}|S_z| \Psi_{\square}}=0$, we now find that
$\braket{\Psi_{\square}|S_z| \Psi_{\square}}=1$. This follows from the fact that
the $e$-excitation at the end of the $S_z$ string are no longer trapped in the
bulk, instead, they lie outside of the bulk region and by using a sequence of
straightforward TN manipulation (cf.~identity (4) in Fig.~\ref{fig:z2_id}), we
can remove them entirely from the system. The intuition behind this construction
is illustrated in Fig.~\ref{fig:zboundary}.

\section{Summary and discussion}\label{sec:sum} In this work, we have considered
three reference systems which are individually of outstanding importance to
condensed matter research: the toric code as an exactly solvable model of long
range entangled matter, the  class D superconductor as an example of a
topological insulator, and the two-dimensional Ising model as a maximally simple
proxy for systems with a discrete symmetry breaking phase transition. These
three system classes are dual to each other. More precisely, the duality
connects the ground states of the TC and the SC with the partition sum of the
IM. Being exact, it extends to all phenomena displayed by the three partner
systems, including their topological or thermal phase transitions, the buildup
of algebraic correlations at criticality, and the presence of topological
boundary modes at domain walls. These equivalences are remarkable in that they
connect phenomena conventionally addressed in different hemispheres of physics
--- such as the phase transition between a spin polarized phase and a
topological spin liquid vs.~the band closing transition between a trivial and a
topological superconductor. 

The dualities discussed in this work are largely known in principle, and have
been derived in previous work by different methods. As they include the duality
between fermionic and bosonic systems, a standard approach is to take a detour
via a transient mapping to one dimensional quantum systems, for which
Bose--Fermi duality is established by Jordan-Wigner transformation. An
alternative approach is to look at them through the coarse graining lens of CFT
and establish equalities between differently realized operators in, say, the
Ising and the free Majorana CFT at criticality. These approaches illustrate the
principle, but arguably lack in the microscopic resolution required when it
comes to the precise comparison of correlation functions. To the best of our
knowledge, this point has first been made in Ref.~\cite{Dotsenko83}, an
observation being that, e.g., the free Majorana correlation function in the SC
becomes that between a spin and a disorder composite operator in the IM where
the exact positioning of the two compound operators on the lattice becomes
crucial. That reference has solved the problem by staying on the two dimensional
lattice and employing (string-) operator algebra to demonstrate that the
spin-disorder operator satisfies the commutation relations of a Majorana
fermion.

In this work, we have proposed an alternative and more comprehensive approach to
the duality, the key idea being to use a \emph{tensor network} as an
intermediate. While at first sight a formulation introducing a fourth player
into a situation that looks complicated already may not look appealing, engaging
a translator TN has various advantages. First, the TN comes in two incarnations
a bosonic and a fermionic one, the passage between the two being explicit, with
no dimensional detours required. Second, both realizations of the TN are
elementary. Being bond dimension two networks, the compound tensors involved in
the construction of the net assume the form of $2\times 2$ matrices, binary
Kronecker $\delta$'s and $\mathbb Z_2$-parity projection tensors. After passing
through a mild learning curve, one can use powerful  graphical relations of
tensor algebra as a resource to obtain results which arguably assume a
substantially more complicated and tedious form in different formulations.
Another attractive feature is that the fermionic TN assumes the form of a
Gaussian Majorana integral. As an alternative to using tensor relations, one may
proceed via techniques otherwise employed in the analysis of free fermion
systems, an approach naturally relevant to the understanding of the
superconductor. In this way, we not only bridge different frameworks within a
single formalism, but also the mindsets of different scientific communities.

While the present work has focused on known manifestations of the duality in the
focus of attention, there are several obvious extensions into less charted
territory. The first is a generalization to non-translationally invariant
systems which, depending on the context, means bit errors, random magnetic
exchange, or static impurities. The tensor network construction has no issues
with the presence of spatially fluctuating bonds, and a natural approach will be
to perform ensemble averaging directly on the level of the TN to generalize the
latter to an effective continuum field theory \cite{Preparation2}. Other
generalizations include the addition of nonlinear contributions to the TN,
asking if they, too, afford a condensed matter interpretation. One may also
study geometric deformations of the TN, for instance the hyperbolic geometries
entering the construction of holographic networks. Here again, it is natural to
ask if and how the holographic bulk boundary correspondence manifests itself in
the dual system classes. 

\section{Acknowledgements}
We acknowledge support from the Deutsche Forschungsgemeinschaft (DFG) under Germany's Excellence Strategy Cluster of Excellence Matter and Light for Quantum Computing (ML4Q) EXC 2004/1 390534769 (A.A.) and within the CRC network TR 183 (project grant 2777101999) as part of projects B04 (A.A and J.E.). It has also received funding from Germany's Excellence Strategy Cluster of Excellence MATH+ (J.E.) and the BMBF (RealistiQ). C.W. acknowledges support from the European Research Council under the European Union Horizon 2020 Research and Innovation Programme via Grant Agreement No.~804213-TMCS.

\bibliography{refs.bib}

\begin{thebibliography}{57}%
\makeatletter
\providecommand \@ifxundefined [1]{%
 \@ifx{#1\undefined}
}%
\providecommand \@ifnum [1]{%
 \ifnum #1\expandafter \@firstoftwo
 \else \expandafter \@secondoftwo
 \fi
}%
\providecommand \@ifx [1]{%
 \ifx #1\expandafter \@firstoftwo
 \else \expandafter \@secondoftwo
 \fi
}%
\providecommand \natexlab [1]{#1}%
\providecommand \enquote  [1]{``#1''}%
\providecommand \bibnamefont  [1]{#1}%
\providecommand \bibfnamefont [1]{#1}%
\providecommand \citenamefont [1]{#1}%
\providecommand \href@noop [0]{\@secondoftwo}%
\providecommand \href [0]{\begingroup \@sanitize@url \@href}%
\providecommand \@href[1]{\@@startlink{#1}\@@href}%
\providecommand \@@href[1]{\endgroup#1\@@endlink}%
\providecommand \@sanitize@url [0]{\catcode `\\12\catcode `\$12\catcode
  `\&12\catcode `\#12\catcode `\^12\catcode `\_12\catcode `\%12\relax}%
\providecommand \@@startlink[1]{}%
\providecommand \@@endlink[0]{}%
\providecommand \url  [0]{\begingroup\@sanitize@url \@url }%
\providecommand \@url [1]{\endgroup\@href {#1}{\urlprefix }}%
\providecommand \urlprefix  [0]{URL }%
\providecommand \Eprint [0]{\href }%
\providecommand \doibase [0]{http://dx.doi.org/}%
\providecommand \selectlanguage [0]{\@gobble}%
\providecommand \bibinfo  [0]{\@secondoftwo}%
\providecommand \bibfield  [0]{\@secondoftwo}%
\providecommand \translation [1]{[#1]}%
\providecommand \BibitemOpen [0]{}%
\providecommand \bibitemStop [0]{}%
\providecommand \bibitemNoStop [0]{.\EOS\space}%
\providecommand \EOS [0]{\spacefactor3000\relax}%
\providecommand \BibitemShut  [1]{\csname bibitem#1\endcsname}%
\let\auto@bib@innerbib\@empty
\bibitem [{\citenamefont {Blackman}(1982)}]{Blackman}%
  \BibitemOpen
  \bibfield  {author} {\bibinfo {author} {\bibfnamefont {J.~A.}\ \bibnamefont
  {Blackman}},\ }\href {\doibase 10.1103/PhysRevB.26.4987} {\bibfield
  {journal} {\bibinfo  {journal} {Phys. Rev. B}\ }\textbf {\bibinfo {volume}
  {26}},\ \bibinfo {pages} {4987} (\bibinfo {year} {1982})}\BibitemShut
  {NoStop}%
\bibitem [{\citenamefont {Blackman}\ and\ \citenamefont
  {Poulter}(1991)}]{PhysRevB.44.4374}%
  \BibitemOpen
  \bibfield  {author} {\bibinfo {author} {\bibfnamefont {J.~A.}\ \bibnamefont
  {Blackman}}\ and\ \bibinfo {author} {\bibfnamefont {J.}~\bibnamefont
  {Poulter}},\ }\href {\doibase 10.1103/PhysRevB.44.4374} {\bibfield  {journal}
  {\bibinfo  {journal} {Phys. Rev. B}\ }\textbf {\bibinfo {volume} {44}},\
  \bibinfo {pages} {4374} (\bibinfo {year} {1991})}\BibitemShut {NoStop}%
\bibitem [{\citenamefont {Blackman}\ \emph {et~al.}(1998)\citenamefont
  {Blackman}, \citenamefont {Gon{\c c}alves},\ and\ \citenamefont
  {Poulter}}]{PhysRevE.58.1502}%
  \BibitemOpen
  \bibfield  {author} {\bibinfo {author} {\bibfnamefont {J.~A.}\ \bibnamefont
  {Blackman}}, \bibinfo {author} {\bibfnamefont {J.~R.}\ \bibnamefont {Gon{\c
  c}alves}}, \ and\ \bibinfo {author} {\bibfnamefont {J.}~\bibnamefont
  {Poulter}},\ }\href {\doibase 10.1103/PhysRevE.58.1502} {\bibfield  {journal}
  {\bibinfo  {journal} {Phys. Rev. E}\ }\textbf {\bibinfo {volume} {58}},\
  \bibinfo {pages} {1502} (\bibinfo {year} {1998})}\BibitemShut {NoStop}%
\bibitem [{\citenamefont {Berezin}(1969)}]{Berezin_1969}%
  \BibitemOpen
  \bibfield  {author} {\bibinfo {author} {\bibfnamefont {F.~A.}\ \bibnamefont
  {Berezin}},\ }\href {\doibase 10.1070/RM1969v024n03ABEH001346} {\bibfield
  {journal} {\bibinfo  {journal} {Russ. Math. Surv.}\ }\textbf {\bibinfo
  {volume} {24}},\ \bibinfo {pages} {1} (\bibinfo {year} {1969})}\BibitemShut
  {NoStop}%
\bibitem [{\citenamefont {Plechko}(1985)}]{PlechkoEnglish}%
  \BibitemOpen
  \bibfield  {author} {\bibinfo {author} {\bibfnamefont {V.~N.}\ \bibnamefont
  {Plechko}},\ }\href {\doibase 10.1007/BF01017042} {\bibfield  {journal}
  {\bibinfo  {journal} {Th. Math. Phys.}\ }\textbf {\bibinfo {volume} {64}},\
  \bibinfo {pages} {748–756} (\bibinfo {year} {1985})}\BibitemShut {NoStop}%
\bibitem [{\citenamefont {Dotsenko}\ and\ \citenamefont
  {Dotsenko}(1983)}]{Dotsenko83}%
  \BibitemOpen
  \bibfield  {author} {\bibinfo {author} {\bibfnamefont {V.~S.}\ \bibnamefont
  {Dotsenko}}\ and\ \bibinfo {author} {\bibfnamefont {V.~S.}\ \bibnamefont
  {Dotsenko}},\ }\href {\doibase 10.1080/00018738300101541} {\bibfield
  {journal} {\bibinfo  {journal} {Adv. Phys.}\ }\textbf {\bibinfo {volume}
  {32}},\ \bibinfo {pages} {129} (\bibinfo {year} {1983})}\BibitemShut
  {NoStop}%
\bibitem [{\citenamefont {Merz}\ and\ \citenamefont
  {Chalker}(2002)}]{ChalkerMerz}%
  \BibitemOpen
  \bibfield  {author} {\bibinfo {author} {\bibfnamefont {F.}~\bibnamefont
  {Merz}}\ and\ \bibinfo {author} {\bibfnamefont {J.~T.}\ \bibnamefont
  {Chalker}},\ }\href {\doibase 10.1103/physrevb.65.054425} {\bibfield
  {journal} {\bibinfo  {journal} {Phys. Rev. B}\ }\textbf {\bibinfo {volume}
  {65}},\ \bibinfo {pages} {054425} (\bibinfo {year} {2002})}\BibitemShut
  {NoStop}%
\bibitem [{\citenamefont {Castelnovo}\ and\ \citenamefont
  {Chamon}(2008)}]{Castelnovo}%
  \BibitemOpen
  \bibfield  {author} {\bibinfo {author} {\bibfnamefont {C.}~\bibnamefont
  {Castelnovo}}\ and\ \bibinfo {author} {\bibfnamefont {C.}~\bibnamefont
  {Chamon}},\ }\href {\doibase 10.1103/physrevb.77.054433} {\bibfield
  {journal} {\bibinfo  {journal} {Phys. Rev. B}\ }\textbf {\bibinfo {volume}
  {77}},\ \bibinfo {pages} {054433} (\bibinfo {year} {2008})}\BibitemShut
  {NoStop}%
\bibitem [{\citenamefont {Zhu}\ and\ \citenamefont {Zhang}(2019)}]{Zhu}%
  \BibitemOpen
  \bibfield  {author} {\bibinfo {author} {\bibfnamefont {G.-Y.}\ \bibnamefont
  {Zhu}}\ and\ \bibinfo {author} {\bibfnamefont {G.-M.}\ \bibnamefont
  {Zhang}},\ }\href {\doibase 10.1103/PhysRevLett.122.176401} {\bibfield
  {journal} {\bibinfo  {journal} {Phys. Rev. Lett.}\ }\textbf {\bibinfo
  {volume} {122}},\ \bibinfo {pages} {176401} (\bibinfo {year}
  {2019})}\BibitemShut {NoStop}%
\bibitem [{\citenamefont {Kitaev}(2003)}]{Kitaev-AnnPhys-2003}%
  \BibitemOpen
  \bibfield  {author} {\bibinfo {author} {\bibfnamefont {A.~Y.}\ \bibnamefont
  {Kitaev}},\ }\href {\doibase http://dx.doi.org/10.1016/S0003-4916(02)00018-0}
  {\bibfield  {journal} {\bibinfo  {journal} {Ann. Phys.}\ }\textbf {\bibinfo
  {volume} {303}},\ \bibinfo {pages} {2 } (\bibinfo {year} {2003})}\BibitemShut
  {NoStop}%
\bibitem [{\citenamefont {Ryu}\ \emph {et~al.}(2010)\citenamefont {Ryu},
  \citenamefont {Schnyder}, \citenamefont {Furusaki},\ and\ \citenamefont
  {Ludwig}}]{Ryu_2010}%
  \BibitemOpen
  \bibfield  {author} {\bibinfo {author} {\bibfnamefont {S.}~\bibnamefont
  {Ryu}}, \bibinfo {author} {\bibfnamefont {A.~P.}\ \bibnamefont {Schnyder}},
  \bibinfo {author} {\bibfnamefont {A.}~\bibnamefont {Furusaki}}, \ and\
  \bibinfo {author} {\bibfnamefont {A.~W.~W.}\ \bibnamefont {Ludwig}},\ }\href
  {\doibase 10.1088/1367-2630/12/6/065010} {\bibfield  {journal} {\bibinfo
  {journal} {New J. Phys.}\ }\textbf {\bibinfo {volume} {12}},\ \bibinfo
  {pages} {065010} (\bibinfo {year} {2010})}\BibitemShut {NoStop}%
\bibitem [{\citenamefont {Onsager}(1944)}]{Onsager}%
  \BibitemOpen
  \bibfield  {author} {\bibinfo {author} {\bibfnamefont {L.}~\bibnamefont
  {Onsager}},\ }\href {\doibase 10.1103/PhysRev.65.117} {\bibfield  {journal}
  {\bibinfo  {journal} {Phys. Rev.}\ }\textbf {\bibinfo {volume} {65}},\
  \bibinfo {pages} {117} (\bibinfo {year} {1944})}\BibitemShut {NoStop}%
\bibitem [{\citenamefont {Kaufman}(1949)}]{Kaufman}%
  \BibitemOpen
  \bibfield  {author} {\bibinfo {author} {\bibfnamefont {B.}~\bibnamefont
  {Kaufman}},\ }\href {\doibase 10.1103/PhysRev.76.1232} {\bibfield  {journal}
  {\bibinfo  {journal} {Phys. Rev.}\ }\textbf {\bibinfo {volume} {76}},\
  \bibinfo {pages} {1232} (\bibinfo {year} {1949})}\BibitemShut {NoStop}%
\bibitem [{\citenamefont {Schultz}\ \emph {et~al.}(1964)\citenamefont
  {Schultz}, \citenamefont {Mattis},\ and\ \citenamefont {Lieb}}]{SCHULTZ}%
  \BibitemOpen
  \bibfield  {author} {\bibinfo {author} {\bibfnamefont {T.~D.}\ \bibnamefont
  {Schultz}}, \bibinfo {author} {\bibfnamefont {D.~C.}\ \bibnamefont {Mattis}},
  \ and\ \bibinfo {author} {\bibfnamefont {E.~H.}\ \bibnamefont {Lieb}},\
  }\href {\doibase 10.1103/RevModPhys.36.856} {\bibfield  {journal} {\bibinfo
  {journal} {Rev. Mod. Phys.}\ }\textbf {\bibinfo {volume} {36}},\ \bibinfo
  {pages} {856} (\bibinfo {year} {1964})}\BibitemShut {NoStop}%
\bibitem [{\citenamefont {Vanhove}\ \emph {et~al.}(2018)\citenamefont
  {Vanhove}, \citenamefont {Bal}, \citenamefont {Williamson}, \citenamefont
  {Bultinck}, \citenamefont {Haegeman},\ and\ \citenamefont
  {Verstraete}}]{Vanhove_2018}%
  \BibitemOpen
  \bibfield  {author} {\bibinfo {author} {\bibfnamefont {R.}~\bibnamefont
  {Vanhove}}, \bibinfo {author} {\bibfnamefont {M.}~\bibnamefont {Bal}},
  \bibinfo {author} {\bibfnamefont {D.~J.}\ \bibnamefont {Williamson}},
  \bibinfo {author} {\bibfnamefont {N.}~\bibnamefont {Bultinck}}, \bibinfo
  {author} {\bibfnamefont {J.}~\bibnamefont {Haegeman}}, \ and\ \bibinfo
  {author} {\bibfnamefont {F.}~\bibnamefont {Verstraete}},\ }\href {\doibase
  10.1103/physrevlett.121.177203} {\bibfield  {journal} {\bibinfo  {journal}
  {Phys. Rev. Lett.}\ }\textbf {\bibinfo {volume} {121}} (\bibinfo {year}
  {2018}),\ 10.1103/physrevlett.121.177203}\BibitemShut {NoStop}%
\bibitem [{\citenamefont {Aasen}\ \emph {et~al.}(2020)\citenamefont {Aasen},
  \citenamefont {Fendley},\ and\ \citenamefont {Mong}}]{aasen2020topological}%
  \BibitemOpen
  \bibfield  {author} {\bibinfo {author} {\bibfnamefont {D.}~\bibnamefont
  {Aasen}}, \bibinfo {author} {\bibfnamefont {P.}~\bibnamefont {Fendley}}, \
  and\ \bibinfo {author} {\bibfnamefont {R.~S.~K.}\ \bibnamefont {Mong}},\
  }\href@noop {} {\enquote {\bibinfo {title} {Topological defects on the
  lattice: Dualities and degeneracies},}\ } (\bibinfo {year} {2020}),\ \Eprint
  {http://arxiv.org/abs/2008.08598} {arXiv:2008.08598 [cond-mat.stat-mech]}
  \BibitemShut {NoStop}%
\bibitem [{\citenamefont {Lootens}\ \emph {et~al.}(2023)\citenamefont
  {Lootens}, \citenamefont {Delcamp}, \citenamefont {Ortiz},\ and\
  \citenamefont {Verstraete}}]{Lootens2023}%
  \BibitemOpen
  \bibfield  {author} {\bibinfo {author} {\bibfnamefont {L.}~\bibnamefont
  {Lootens}}, \bibinfo {author} {\bibfnamefont {C.}~\bibnamefont {Delcamp}},
  \bibinfo {author} {\bibfnamefont {G.}~\bibnamefont {Ortiz}}, \ and\ \bibinfo
  {author} {\bibfnamefont {F.}~\bibnamefont {Verstraete}},\ }\href {\doibase
  10.1103/prxquantum.4.020357} {\bibfield  {journal} {\bibinfo  {journal} {PRX
  Quantum}\ }\textbf {\bibinfo {volume} {4}} (\bibinfo {year} {2023}),\
  10.1103/prxquantum.4.020357}\BibitemShut {NoStop}%
\bibitem [{\citenamefont {Lootens}\ \emph {et~al.}(2022)\citenamefont
  {Lootens}, \citenamefont {Delcamp},\ and\ \citenamefont
  {Verstraete}}]{lootens2022dualities}%
  \BibitemOpen
  \bibfield  {author} {\bibinfo {author} {\bibfnamefont {L.}~\bibnamefont
  {Lootens}}, \bibinfo {author} {\bibfnamefont {C.}~\bibnamefont {Delcamp}}, \
  and\ \bibinfo {author} {\bibfnamefont {F.}~\bibnamefont {Verstraete}},\
  }\href@noop {} {\enquote {\bibinfo {title} {Dualities in one-dimensional
  quantum lattice models: topological sectors},}\ } (\bibinfo {year} {2022}),\
  \Eprint {http://arxiv.org/abs/2211.03777} {arXiv:2211.03777 [quant-ph]}
  \BibitemShut {NoStop}%
\bibitem [{\citenamefont {Valiant}(2002)}]{Valiant}%
  \BibitemOpen
  \bibfield  {author} {\bibinfo {author} {\bibfnamefont {L.~G.}\ \bibnamefont
  {Valiant}},\ }\href {\doibase 10.1137/S0097539700377025} {\bibfield
  {journal} {\bibinfo  {journal} {SIAM J. Comp.}\ }\textbf {\bibinfo {volume}
  {31}},\ \bibinfo {pages} {1229} (\bibinfo {year} {2002})}\BibitemShut
  {NoStop}%
\bibitem [{\citenamefont {Bravyi}(2009)}]{Bravyi2009}%
  \BibitemOpen
  \bibfield  {author} {\bibinfo {author} {\bibfnamefont {S.}~\bibnamefont
  {Bravyi}},\ }\href {\doibase 10.1090/conm/482/09419} {\bibfield  {journal}
  {\bibinfo  {journal} {Adv. Quant. Comp.}\ }\textbf {\bibinfo {volume}
  {482}},\ \bibinfo {pages} {179} (\bibinfo {year} {2009})}\BibitemShut
  {NoStop}%
\bibitem [{\citenamefont {Jahn}\ \emph {et~al.}(2019)\citenamefont {Jahn},
  \citenamefont {Gluza}, \citenamefont {Pastawski},\ and\ \citenamefont
  {Eisert}}]{Jahn}%
  \BibitemOpen
  \bibfield  {author} {\bibinfo {author} {\bibfnamefont {A.}~\bibnamefont
  {Jahn}}, \bibinfo {author} {\bibfnamefont {M.}~\bibnamefont {Gluza}},
  \bibinfo {author} {\bibfnamefont {F.}~\bibnamefont {Pastawski}}, \ and\
  \bibinfo {author} {\bibfnamefont {J.}~\bibnamefont {Eisert}},\ }\href
  {\doibase 10.1126/sciadv.aaw0092} {\bibfield  {journal} {\bibinfo  {journal}
  {Science Adv.}\ }\textbf {\bibinfo {volume} {5}},\ \bibinfo {pages}
  {eaaw0092} (\bibinfo {year} {2019})}\BibitemShut {NoStop}%
\bibitem [{\citenamefont {Cai}\ and\ \citenamefont {Choudhary}(2006)}]{Cai}%
  \BibitemOpen
  \bibfield  {author} {\bibinfo {author} {\bibfnamefont {J.-Y.}\ \bibnamefont
  {Cai}}\ and\ \bibinfo {author} {\bibfnamefont {V.}~\bibnamefont
  {Choudhary}},\ }in\ \href@noop {} {\emph {\bibinfo {booktitle} {Theory and
  Applications of Models of Computation}}},\ \bibinfo {editor} {edited by\
  \bibinfo {editor} {\bibfnamefont {J.-Y.}\ \bibnamefont {Cai}}, \bibinfo
  {editor} {\bibfnamefont {S.~B.}\ \bibnamefont {Cooper}}, \ and\ \bibinfo
  {editor} {\bibfnamefont {A.}~\bibnamefont {Li}}}\ (\bibinfo  {publisher}
  {Springer Berlin Heidelberg},\ \bibinfo {address} {Berlin, Heidelberg},\
  \bibinfo {year} {2006})\ pp.\ \bibinfo {pages} {248--261}\BibitemShut
  {NoStop}%
\bibitem [{\citenamefont {Cai}\ \emph {et~al.}(2007)\citenamefont {Cai},
  \citenamefont {Choudhary},\ and\ \citenamefont {Lu}}]{Cai2007OnTT}%
  \BibitemOpen
  \bibfield  {author} {\bibinfo {author} {\bibfnamefont {J.-Y.}\ \bibnamefont
  {Cai}}, \bibinfo {author} {\bibfnamefont {V.}~\bibnamefont {Choudhary}}, \
  and\ \bibinfo {author} {\bibfnamefont {P.}~\bibnamefont {Lu}},\ }\href
  {\doibase 10.1007/s00224-007-9092-8} {\bibfield  {journal} {\bibinfo
  {journal} {Th. Comp. Syst.}\ }\textbf {\bibinfo {volume} {45}},\ \bibinfo
  {pages} {108} (\bibinfo {year} {2007})}\BibitemShut {NoStop}%
\bibitem [{\citenamefont {Derby}\ \emph {et~al.}(2021)\citenamefont {Derby},
  \citenamefont {Klassen}, \citenamefont {Bausch},\ and\ \citenamefont
  {Cubitt}}]{PhysRevB.104.035118}%
  \BibitemOpen
  \bibfield  {author} {\bibinfo {author} {\bibfnamefont {C.}~\bibnamefont
  {Derby}}, \bibinfo {author} {\bibfnamefont {J.}~\bibnamefont {Klassen}},
  \bibinfo {author} {\bibfnamefont {J.}~\bibnamefont {Bausch}}, \ and\ \bibinfo
  {author} {\bibfnamefont {T.}~\bibnamefont {Cubitt}},\ }\href {\doibase
  10.1103/PhysRevB.104.035118} {\bibfield  {journal} {\bibinfo  {journal}
  {Phys. Rev. B}\ }\textbf {\bibinfo {volume} {104}},\ \bibinfo {pages}
  {035118} (\bibinfo {year} {2021})}\BibitemShut {NoStop}%
\bibitem [{\citenamefont {Kadanoff}\ and\ \citenamefont
  {Ceva}(1971)}]{KadanoffCeva}%
  \BibitemOpen
  \bibfield  {author} {\bibinfo {author} {\bibfnamefont {L.~P.}\ \bibnamefont
  {Kadanoff}}\ and\ \bibinfo {author} {\bibfnamefont {H.}~\bibnamefont
  {Ceva}},\ }\href {\doibase 10.1103/PhysRevB.3.3918} {\bibfield  {journal}
  {\bibinfo  {journal} {Phys. Rev. B}\ }\textbf {\bibinfo {volume} {3}},\
  \bibinfo {pages} {3918} (\bibinfo {year} {1971})}\BibitemShut {NoStop}%
\bibitem [{\citenamefont {Fradkin}(2017)}]{Fradkin_2017}%
  \BibitemOpen
  \bibfield  {author} {\bibinfo {author} {\bibfnamefont {E.}~\bibnamefont
  {Fradkin}},\ }\href {\doibase 10.1007/s10955-017-1737-7} {\bibfield
  {journal} {\bibinfo  {journal} {Journal of Statistical Physics}\ }\textbf
  {\bibinfo {volume} {167}},\ \bibinfo {pages} {427} (\bibinfo {year}
  {2017})}\BibitemShut {NoStop}%
\bibitem [{\citenamefont {Cirac}\ \emph {et~al.}(2021)\citenamefont {Cirac},
  \citenamefont {Perez-Garcia}, \citenamefont {Schuch},\ and\ \citenamefont
  {Verstraete}}]{Cirac21}%
  \BibitemOpen
  \bibfield  {author} {\bibinfo {author} {\bibfnamefont {J.~I.}\ \bibnamefont
  {Cirac}}, \bibinfo {author} {\bibfnamefont {D.}~\bibnamefont {Perez-Garcia}},
  \bibinfo {author} {\bibfnamefont {N.}~\bibnamefont {Schuch}}, \ and\ \bibinfo
  {author} {\bibfnamefont {F.}~\bibnamefont {Verstraete}},\ }\href {\doibase
  10.1103/RevModPhys.93.045003} {\bibfield  {journal} {\bibinfo  {journal}
  {Rev. Mod. Phys.}\ }\textbf {\bibinfo {volume} {93}},\ \bibinfo {pages}
  {045003} (\bibinfo {year} {2021})}\BibitemShut {NoStop}%
\bibitem [{\citenamefont {Wille}\ \emph
  {et~al.}(2023{\natexlab{a}})\citenamefont {Wille}, \citenamefont {Eisert},\
  and\ \citenamefont {Altland}}]{Preparation}%
  \BibitemOpen
  \bibfield  {author} {\bibinfo {author} {\bibfnamefont {C.}~\bibnamefont
  {Wille}}, \bibinfo {author} {\bibfnamefont {J.}~\bibnamefont {Eisert}}, \
  and\ \bibinfo {author} {\bibfnamefont {A.}~\bibnamefont {Altland}},\
  }\href@noop {} {\enquote {\bibinfo {title} {In preparation},}\ } (\bibinfo
  {year} {2023}{\natexlab{a}})\BibitemShut {NoStop}%
\bibitem [{\citenamefont {Wille}\ \emph
  {et~al.}(2023{\natexlab{b}})\citenamefont {Wille}, \citenamefont {Eisert},
  \citenamefont {Jahn},\ and\ \citenamefont {Altland}}]{Preparation2}%
  \BibitemOpen
  \bibfield  {author} {\bibinfo {author} {\bibfnamefont {C.}~\bibnamefont
  {Wille}}, \bibinfo {author} {\bibfnamefont {J.}~\bibnamefont {Eisert}},
  \bibinfo {author} {\bibfnamefont {A.}~\bibnamefont {Jahn}}, \ and\ \bibinfo
  {author} {\bibfnamefont {A.}~\bibnamefont {Altland}},\ }\href@noop {}
  {\enquote {\bibinfo {title} {In preparation},}\ } (\bibinfo {year}
  {2023}{\natexlab{b}})\BibitemShut {NoStop}%
\bibitem [{\citenamefont {Or\'us}(2014)}]{Orus-AnnPhys-2014}%
  \BibitemOpen
  \bibfield  {author} {\bibinfo {author} {\bibfnamefont {R.}~\bibnamefont
  {Or\'us}},\ }\href {\doibase 10.1016/j.aop.2014.06.013} {\bibfield  {journal}
  {\bibinfo  {journal} {Ann. Phys.}\ }\textbf {\bibinfo {volume} {349}},\
  \bibinfo {pages} {117} (\bibinfo {year} {2014})}\BibitemShut {NoStop}%
\bibitem [{\citenamefont {Eisert}\ \emph {et~al.}(2010)\citenamefont {Eisert},
  \citenamefont {Cramer},\ and\ \citenamefont {Plenio}}]{AreaReview}%
  \BibitemOpen
  \bibfield  {author} {\bibinfo {author} {\bibfnamefont {J.}~\bibnamefont
  {Eisert}}, \bibinfo {author} {\bibfnamefont {M.}~\bibnamefont {Cramer}}, \
  and\ \bibinfo {author} {\bibfnamefont {M.~B.}\ \bibnamefont {Plenio}},\
  }\href {\doibase 10.1103/RevModPhys.82.277} {\bibfield  {journal} {\bibinfo
  {journal} {Rev. Mod. Phys.}\ }\textbf {\bibinfo {volume} {82}},\ \bibinfo
  {pages} {277} (\bibinfo {year} {2010})}\BibitemShut {NoStop}%
\bibitem [{\citenamefont {Bridgeman}\ and\ \citenamefont
  {Chubb}(2017)}]{Bridgeman2017}%
  \BibitemOpen
  \bibfield  {author} {\bibinfo {author} {\bibfnamefont {J.~C.}\ \bibnamefont
  {Bridgeman}}\ and\ \bibinfo {author} {\bibfnamefont {C.~T.}\ \bibnamefont
  {Chubb}},\ }\href {\doibase 10.1088/1751-8121/aa6dc3} {\bibfield  {journal}
  {\bibinfo  {journal} {J. Phys. A}\ }\textbf {\bibinfo {volume} {50}},\
  \bibinfo {pages} {223001} (\bibinfo {year} {2017})}\BibitemShut {NoStop}%
\bibitem [{\citenamefont {Or{\'u}s}(2019)}]{Orus2019}%
  \BibitemOpen
  \bibfield  {author} {\bibinfo {author} {\bibfnamefont {R.}~\bibnamefont
  {Or{\'u}s}},\ }\href {\doibase 10.1038/s42254-019-0086-7} {\bibfield
  {journal} {\bibinfo  {journal} {Nature Rev. Phys.}\ }\textbf {\bibinfo
  {volume} {1}},\ \bibinfo {pages} {538} (\bibinfo {year} {2019})}\BibitemShut
  {NoStop}%
\bibitem [{\citenamefont {Barthel}\ \emph {et~al.}(2009)\citenamefont
  {Barthel}, \citenamefont {Pineda},\ and\ \citenamefont
  {Eisert}}]{PhysRevA.80.042333}%
  \BibitemOpen
  \bibfield  {author} {\bibinfo {author} {\bibfnamefont {T.}~\bibnamefont
  {Barthel}}, \bibinfo {author} {\bibfnamefont {C.}~\bibnamefont {Pineda}}, \
  and\ \bibinfo {author} {\bibfnamefont {J.}~\bibnamefont {Eisert}},\ }\href
  {\doibase 10.1103/PhysRevA.80.042333} {\bibfield  {journal} {\bibinfo
  {journal} {Phys. Rev. A}\ }\textbf {\bibinfo {volume} {80}},\ \bibinfo
  {pages} {042333} (\bibinfo {year} {2009})}\BibitemShut {NoStop}%
\bibitem [{\citenamefont {Corboz}\ \emph {et~al.}(2010)\citenamefont {Corboz},
  \citenamefont {Orus}, \citenamefont {Bauer},\ and\ \citenamefont
  {Vidal}}]{CorbozPEPSFermions}%
  \BibitemOpen
  \bibfield  {author} {\bibinfo {author} {\bibfnamefont {P.}~\bibnamefont
  {Corboz}}, \bibinfo {author} {\bibfnamefont {R.}~\bibnamefont {Orus}},
  \bibinfo {author} {\bibfnamefont {B.}~\bibnamefont {Bauer}}, \ and\ \bibinfo
  {author} {\bibfnamefont {G.}~\bibnamefont {Vidal}},\ }\href {\doibase
  10.1103/PhysRevB.81.165104} {\bibfield  {journal} {\bibinfo  {journal} {Phys.
  Rev. B}\ }\textbf {\bibinfo {volume} {81}},\ \bibinfo {pages} {165104}
  (\bibinfo {year} {2010})}\BibitemShut {NoStop}%
\bibitem [{\citenamefont {Pi\ifmmode~\check{z}\else \v{z}\fi{}orn}\ and\
  \citenamefont {Verstraete}(2010)}]{PhysRevB.81.245110}%
  \BibitemOpen
  \bibfield  {author} {\bibinfo {author} {\bibfnamefont {I.}~\bibnamefont
  {Pi\ifmmode~\check{z}\else \v{z}\fi{}orn}}\ and\ \bibinfo {author}
  {\bibfnamefont {F.}~\bibnamefont {Verstraete}},\ }\href {\doibase
  10.1103/PhysRevB.81.245110} {\bibfield  {journal} {\bibinfo  {journal} {Phys.
  Rev. B}\ }\textbf {\bibinfo {volume} {81}},\ \bibinfo {pages} {245110}
  (\bibinfo {year} {2010})}\BibitemShut {NoStop}%
\bibitem [{\citenamefont {Kraus}\ \emph {et~al.}(2010)\citenamefont {Kraus},
  \citenamefont {Schuch}, \citenamefont {Verstraete},\ and\ \citenamefont
  {Cirac}}]{Kraus}%
  \BibitemOpen
  \bibfield  {author} {\bibinfo {author} {\bibfnamefont {C.~V.}\ \bibnamefont
  {Kraus}}, \bibinfo {author} {\bibfnamefont {N.}~\bibnamefont {Schuch}},
  \bibinfo {author} {\bibfnamefont {F.}~\bibnamefont {Verstraete}}, \ and\
  \bibinfo {author} {\bibfnamefont {J.~I.}\ \bibnamefont {Cirac}},\ }\href
  {\doibase 10.1103/physreva.81.052338} {\bibfield  {journal} {\bibinfo
  {journal} {Phys. Rev. A}\ }\textbf {\bibinfo {volume} {81}},\ \bibinfo
  {pages} {052338} (\bibinfo {year} {2010})}\BibitemShut {NoStop}%
\bibitem [{\citenamefont {Gu}\ \emph {et~al.}(2010)\citenamefont {Gu},
  \citenamefont {Verstraete},\ and\ \citenamefont {Wen}}]{gu2010grassmann}%
  \BibitemOpen
  \bibfield  {author} {\bibinfo {author} {\bibfnamefont {Z.-C.}\ \bibnamefont
  {Gu}}, \bibinfo {author} {\bibfnamefont {F.}~\bibnamefont {Verstraete}}, \
  and\ \bibinfo {author} {\bibfnamefont {X.-G.}\ \bibnamefont {Wen}},\
  }\href@noop {} {\enquote {\bibinfo {title} {Grassmann tensor network states
  and its renormalization for strongly correlated fermionic and bosonic
  states},}\ } (\bibinfo {year} {2010}),\ \Eprint
  {http://arxiv.org/abs/1004.2563} {arXiv:1004.2563} \BibitemShut {NoStop}%
\bibitem [{\citenamefont {Bultinck}\ \emph {et~al.}(2017)\citenamefont
  {Bultinck}, \citenamefont {Williamson}, \citenamefont {Haegeman},\ and\
  \citenamefont {Verstraete}}]{Nick}%
  \BibitemOpen
  \bibfield  {author} {\bibinfo {author} {\bibfnamefont {N.}~\bibnamefont
  {Bultinck}}, \bibinfo {author} {\bibfnamefont {D.~J.}\ \bibnamefont
  {Williamson}}, \bibinfo {author} {\bibfnamefont {J.}~\bibnamefont
  {Haegeman}}, \ and\ \bibinfo {author} {\bibfnamefont {F.}~\bibnamefont
  {Verstraete}},\ }\href {\doibase 10.1088/1751-8121/aa99cc} {\bibfield
  {journal} {\bibinfo  {journal} {J. Phys. A}\ }\textbf {\bibinfo {volume}
  {51}},\ \bibinfo {pages} {025202} (\bibinfo {year} {2017})}\BibitemShut
  {NoStop}%
\bibitem [{\citenamefont {Wille}\ \emph {et~al.}(2017)\citenamefont {Wille},
  \citenamefont {Buerschaper},\ and\ \citenamefont
  {Eisert}}]{PhysRevB.95.245127}%
  \BibitemOpen
  \bibfield  {author} {\bibinfo {author} {\bibfnamefont {C.}~\bibnamefont
  {Wille}}, \bibinfo {author} {\bibfnamefont {O.}~\bibnamefont {Buerschaper}},
  \ and\ \bibinfo {author} {\bibfnamefont {J.}~\bibnamefont {Eisert}},\ }\href
  {\doibase 10.1103/PhysRevB.95.245127} {\bibfield  {journal} {\bibinfo
  {journal} {Phys. Rev. B}\ }\textbf {\bibinfo {volume} {95}},\ \bibinfo
  {pages} {245127} (\bibinfo {year} {2017})}\BibitemShut {NoStop}%
\bibitem [{Note1()}]{Note1}%
  \BibitemOpen
  \bibinfo {note} {For example, for tensors with four indices, we observe that
  a matchgate tensor has six free parameters, whereas a weighted $\protect
  \mathbb Z_2$ tensor has only four.}\BibitemShut {Stop}%
\bibitem [{\citenamefont {Gogolin}\ \emph {et~al.}(2004)\citenamefont
  {Gogolin}, \citenamefont {Nersesyan},\ and\ \citenamefont
  {Tsvelik}}]{GogolinBook2004}%
  \BibitemOpen
  \bibfield  {author} {\bibinfo {author} {\bibfnamefont {A.~O.}\ \bibnamefont
  {Gogolin}}, \bibinfo {author} {\bibfnamefont {A.~A.}\ \bibnamefont
  {Nersesyan}}, \ and\ \bibinfo {author} {\bibfnamefont {A.~M.}\ \bibnamefont
  {Tsvelik}},\ }\href@noop {} {\emph {\bibinfo {title} {Bosonization and
  strongly correlated systems}}}\ (\bibinfo  {publisher} {Cambridge University
  Press},\ \bibinfo {year} {2004})\BibitemShut {NoStop}%
\bibitem [{\citenamefont {Kac}\ and\ \citenamefont {Ward}(1952)}]{KacWard}%
  \BibitemOpen
  \bibfield  {author} {\bibinfo {author} {\bibfnamefont {M.}~\bibnamefont
  {Kac}}\ and\ \bibinfo {author} {\bibfnamefont {J.~C.}\ \bibnamefont {Ward}},\
  }\href {\doibase 10.1103/PhysRev.88.1332} {\bibfield  {journal} {\bibinfo
  {journal} {Phys. Rev.}\ }\textbf {\bibinfo {volume} {88}},\ \bibinfo {pages}
  {1332} (\bibinfo {year} {1952})}\BibitemShut {NoStop}%
\bibitem [{\citenamefont {Hurst}\ and\ \citenamefont
  {Green}(2004)}]{HurstGreen}%
  \BibitemOpen
  \bibfield  {author} {\bibinfo {author} {\bibfnamefont {C.~A.}\ \bibnamefont
  {Hurst}}\ and\ \bibinfo {author} {\bibfnamefont {H.~S.}\ \bibnamefont
  {Green}},\ }\href {\doibase 10.1063/1.1731333} {\bibfield  {journal}
  {\bibinfo  {journal} {J. Chem. Phys.}\ }\textbf {\bibinfo {volume} {33}},\
  \bibinfo {pages} {1059} (\bibinfo {year} {2004})}\BibitemShut {NoStop}%
\bibitem [{\citenamefont {Cho}\ and\ \citenamefont {Fisher}(1997)}]{Cho1997}%
  \BibitemOpen
  \bibfield  {author} {\bibinfo {author} {\bibfnamefont {S.}~\bibnamefont
  {Cho}}\ and\ \bibinfo {author} {\bibfnamefont {M.~P.~A.}\ \bibnamefont
  {Fisher}},\ }\href {\doibase 10.1103/physrevb.55.1025} {\bibfield  {journal}
  {\bibinfo  {journal} {Physical Review B}\ }\textbf {\bibinfo {volume} {55}},\
  \bibinfo {pages} {1025} (\bibinfo {year} {1997})}\BibitemShut {NoStop}%
\bibitem [{\citenamefont {Read}\ and\ \citenamefont
  {Ludwig}(2000)}]{Read_2000}%
  \BibitemOpen
  \bibfield  {author} {\bibinfo {author} {\bibfnamefont {N.}~\bibnamefont
  {Read}}\ and\ \bibinfo {author} {\bibfnamefont {A.~W.~W.}\ \bibnamefont
  {Ludwig}},\ }\href {\doibase 10.1103/physrevb.63.024404} {\bibfield
  {journal} {\bibinfo  {journal} {Physical Review B}\ }\textbf {\bibinfo
  {volume} {63}} (\bibinfo {year} {2000}),\
  10.1103/physrevb.63.024404}\BibitemShut {NoStop}%
\bibitem [{\citenamefont {Gruzberg}\ \emph {et~al.}(2001)\citenamefont
  {Gruzberg}, \citenamefont {Read},\ and\ \citenamefont {Ludwig}}]{Gruzberg}%
  \BibitemOpen
  \bibfield  {author} {\bibinfo {author} {\bibfnamefont {I.~A.}\ \bibnamefont
  {Gruzberg}}, \bibinfo {author} {\bibfnamefont {N.}~\bibnamefont {Read}}, \
  and\ \bibinfo {author} {\bibfnamefont {A.~W.~W.}\ \bibnamefont {Ludwig}},\
  }\href {\doibase 10.1103/PhysRevB.63.104422} {\bibfield  {journal} {\bibinfo
  {journal} {Phys. Rev. B}\ }\textbf {\bibinfo {volume} {63}},\ \bibinfo
  {pages} {104422} (\bibinfo {year} {2001})}\BibitemShut {NoStop}%
\bibitem [{\citenamefont {Jian}\ \emph {et~al.}(2022)\citenamefont {Jian},
  \citenamefont {Bauer}, \citenamefont {Keselman},\ and\ \citenamefont
  {Ludwig}}]{Ludwig22}%
  \BibitemOpen
  \bibfield  {author} {\bibinfo {author} {\bibfnamefont {C.-M.}\ \bibnamefont
  {Jian}}, \bibinfo {author} {\bibfnamefont {B.}~\bibnamefont {Bauer}},
  \bibinfo {author} {\bibfnamefont {A.}~\bibnamefont {Keselman}}, \ and\
  \bibinfo {author} {\bibfnamefont {A.~W.~W.}\ \bibnamefont {Ludwig}},\ }\href
  {\doibase 10.1103/PhysRevB.106.134206} {\bibfield  {journal} {\bibinfo
  {journal} {Phys. Rev. B}\ }\textbf {\bibinfo {volume} {106}},\ \bibinfo
  {pages} {134206} (\bibinfo {year} {2022})}\BibitemShut {NoStop}%
\bibitem [{Note2()}]{Note2}%
  \BibitemOpen
  \bibinfo {note} {We multiply each link by $1=\protect \sqrt {2} / \protect
  \sqrt {2}$, where the $\protect \sqrt {2}$ in the numerator multiplies our
  $T$ and the ones in the denominator remove the factor of $2$ multiplying the
  central $\delta $-tensors.}\BibitemShut {Stop}%
\bibitem [{\citenamefont {Isakov}\ \emph {et~al.}(2011)\citenamefont {Isakov},
  \citenamefont {Fendley}, \citenamefont {Ludwig}, \citenamefont {Trebst},\
  and\ \citenamefont {Troyer}}]{Isakov}%
  \BibitemOpen
  \bibfield  {author} {\bibinfo {author} {\bibfnamefont {S.~V.}\ \bibnamefont
  {Isakov}}, \bibinfo {author} {\bibfnamefont {P.}~\bibnamefont {Fendley}},
  \bibinfo {author} {\bibfnamefont {A.~W.~W.}\ \bibnamefont {Ludwig}}, \bibinfo
  {author} {\bibfnamefont {S.}~\bibnamefont {Trebst}}, \ and\ \bibinfo {author}
  {\bibfnamefont {M.}~\bibnamefont {Troyer}},\ }\href {\doibase
  10.1103/PhysRevB.83.125114} {\bibfield  {journal} {\bibinfo  {journal} {Phys.
  Rev. B}\ }\textbf {\bibinfo {volume} {83}},\ \bibinfo {pages} {125114}
  (\bibinfo {year} {2011})}\BibitemShut {NoStop}%
\bibitem [{\citenamefont {Dennis}\ \emph {et~al.}(2002)\citenamefont {Dennis},
  \citenamefont {Kitaev}, \citenamefont {Landahl},\ and\ \citenamefont
  {Preskill}}]{TopologicalQuantumMemory}%
  \BibitemOpen
  \bibfield  {author} {\bibinfo {author} {\bibfnamefont {E.}~\bibnamefont
  {Dennis}}, \bibinfo {author} {\bibfnamefont {A.}~\bibnamefont {Kitaev}},
  \bibinfo {author} {\bibfnamefont {A.}~\bibnamefont {Landahl}}, \ and\
  \bibinfo {author} {\bibfnamefont {J.}~\bibnamefont {Preskill}},\ }\href
  {\doibase 10.1063/1.1499754} {\bibfield  {journal} {\bibinfo  {journal} {J.
  Math. Phys.}\ }\textbf {\bibinfo {volume} {43}},\ \bibinfo {pages} {4452}
  (\bibinfo {year} {2002})}\BibitemShut {NoStop}%
\bibitem [{\citenamefont {Verstraete}\ \emph {et~al.}(2006)\citenamefont
  {Verstraete}, \citenamefont {Wolf}, \citenamefont {Perez-Garcia},\ and\
  \citenamefont {Cirac}}]{Verstraete_2006}%
  \BibitemOpen
  \bibfield  {author} {\bibinfo {author} {\bibfnamefont {F.}~\bibnamefont
  {Verstraete}}, \bibinfo {author} {\bibfnamefont {M.~M.}\ \bibnamefont
  {Wolf}}, \bibinfo {author} {\bibfnamefont {D.}~\bibnamefont {Perez-Garcia}},
  \ and\ \bibinfo {author} {\bibfnamefont {J.~I.}\ \bibnamefont {Cirac}},\
  }\href {\doibase 10.1103/physrevlett.96.220601} {\bibfield  {journal}
  {\bibinfo  {journal} {Phys. Rev. Lett.}\ }\textbf {\bibinfo {volume} {96}},\
  \bibinfo {pages} {220601} (\bibinfo {year} {2006})}\BibitemShut {NoStop}%
\bibitem [{\citenamefont {Gu}\ \emph {et~al.}(2009)\citenamefont {Gu},
  \citenamefont {Levin}, \citenamefont {Swingle},\ and\ \citenamefont
  {Wen}}]{Gu_2009}%
  \BibitemOpen
  \bibfield  {author} {\bibinfo {author} {\bibfnamefont {Z.-C.}\ \bibnamefont
  {Gu}}, \bibinfo {author} {\bibfnamefont {M.}~\bibnamefont {Levin}}, \bibinfo
  {author} {\bibfnamefont {B.}~\bibnamefont {Swingle}}, \ and\ \bibinfo
  {author} {\bibfnamefont {X.-G.}\ \bibnamefont {Wen}},\ }\href {\doibase
  10.1103/physrevb.79.085118} {\bibfield  {journal} {\bibinfo  {journal} {Phys.
  Rev. B}\ }\textbf {\bibinfo {volume} {79}},\ \bibinfo {pages} {085118}
  (\bibinfo {year} {2009})}\BibitemShut {NoStop}%
\bibitem [{\citenamefont {Schuch}\ \emph {et~al.}(2010)\citenamefont {Schuch},
  \citenamefont {Cirac},\ and\ \citenamefont {Perez-Garcia}}]{PEPSTopology}%
  \BibitemOpen
  \bibfield  {author} {\bibinfo {author} {\bibfnamefont {N.}~\bibnamefont
  {Schuch}}, \bibinfo {author} {\bibfnamefont {I.}~\bibnamefont {Cirac}}, \
  and\ \bibinfo {author} {\bibfnamefont {D.}~\bibnamefont {Perez-Garcia}},\
  }\href {\doibase 10.1016/j.aop.2010.05.008} {\bibfield  {journal} {\bibinfo
  {journal} {Ann. Phys.}\ }\textbf {\bibinfo {volume} {325}},\ \bibinfo {pages}
  {2153} (\bibinfo {year} {2010})}\BibitemShut {NoStop}%
\bibitem [{\citenamefont {Haldane}(1988)}]{PhysRevLett.61.2015}%
  \BibitemOpen
  \bibfield  {author} {\bibinfo {author} {\bibfnamefont {F.~D.~M.}\
  \bibnamefont {Haldane}},\ }\href {\doibase 10.1103/PhysRevLett.61.2015}
  {\bibfield  {journal} {\bibinfo  {journal} {Phys. Rev. Lett.}\ }\textbf
  {\bibinfo {volume} {61}},\ \bibinfo {pages} {2015} (\bibinfo {year}
  {1988})}\BibitemShut {NoStop}%
\bibitem [{\citenamefont {Fu}\ and\ \citenamefont {Kane}(2008)}]{FuKane2008}%
  \BibitemOpen
  \bibfield  {author} {\bibinfo {author} {\bibfnamefont {L.}~\bibnamefont
  {Fu}}\ and\ \bibinfo {author} {\bibfnamefont {C.~L.}\ \bibnamefont {Kane}},\
  }\href {\doibase 10.1103/PhysRevLett.100.096407} {\bibfield  {journal}
  {\bibinfo  {journal} {Phys. Rev. Lett.}\ }\textbf {\bibinfo {volume} {100}},\
  \bibinfo {pages} {096407} (\bibinfo {year} {2008})}\BibitemShut {NoStop}%
\bibitem [{Note3()}]{Note3}%
  \BibitemOpen
  \bibinfo {note} {By contrast, the ground state of a TC with \protect \emph
  {rough boundaries} where bonds poke into the vacuum contains strings ending
  there.}\BibitemShut {Stop}%
\end{thebibliography}%


\clearpage

\appendix

\section{Mapping bosonic to fermionic tensor networks}\label{app:bosonfermion}
\begin{figure}
	\centering
	\includegraphics[width=0.45\textwidth]{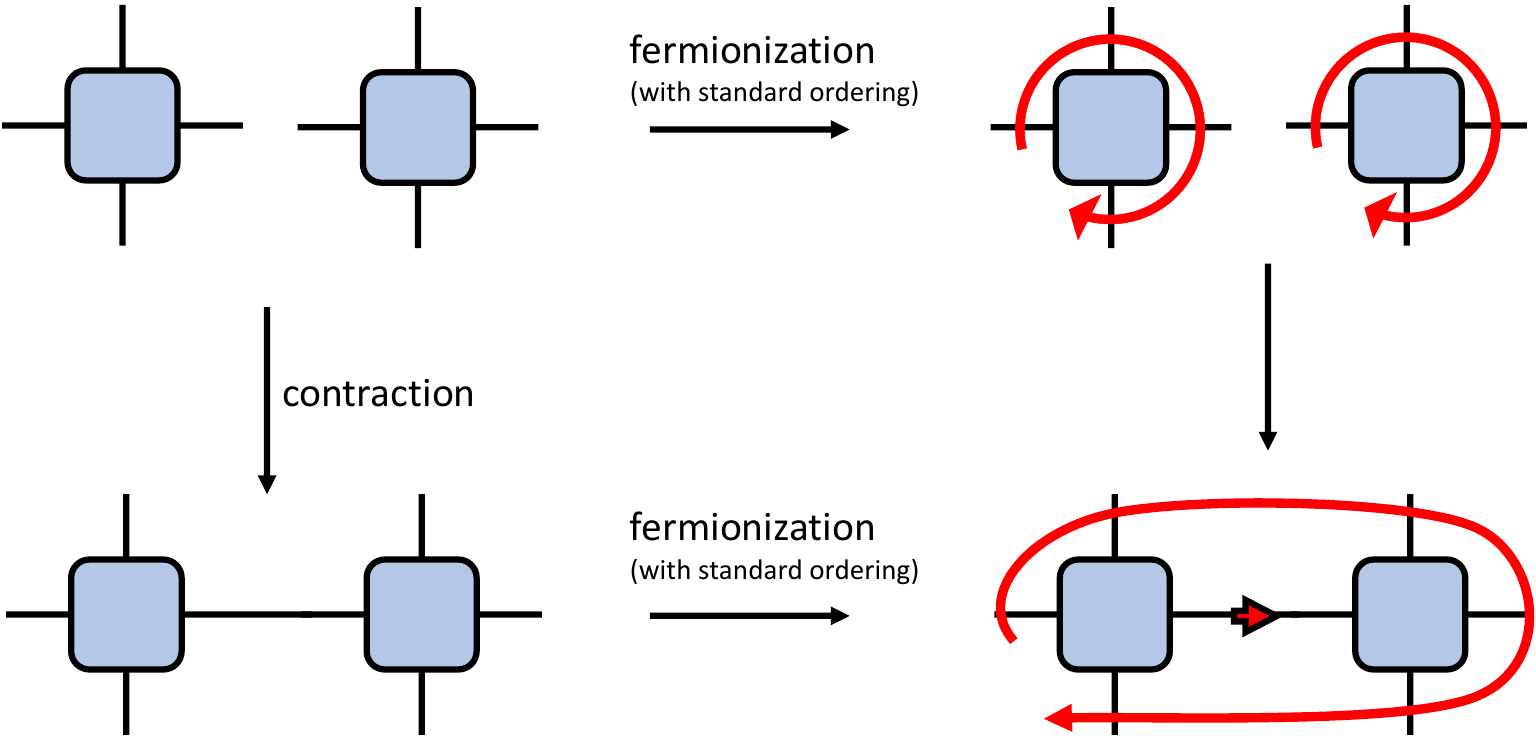}
	\caption{Tensor contraction versus fermionization. The diagram is non-commutative in general. During the fermionization process, we need to choose an ordering of the fermionic modes. We always choose so-called standard ordering defined in the main text. For the fermionic contraction, we need to choose a bond orientation and again choose a standard orientation.}
	\label{fig:diagram}
\end{figure}

In the following, we discuss how the map from bosonic to fermionic tensors
behaves under tensor contraction. In particular, we investigate under which
conditions the diagram in Fig.~\ref{fig:diagram} is commutative or almost
commutative in a sense to be specified momentarily. We recapitulate that we need
to choose an orientation for the contraction of the fermionic bonds and an
ordering of the contracted bosonic tensor for their fermionization. We will
focus on rectangular tensor network patches on a square lattice and show that a
particular choice of orientations and orderings yields consistent results. 

\begin{figure}
	\centering
	\includegraphics[width=0.5\textwidth]{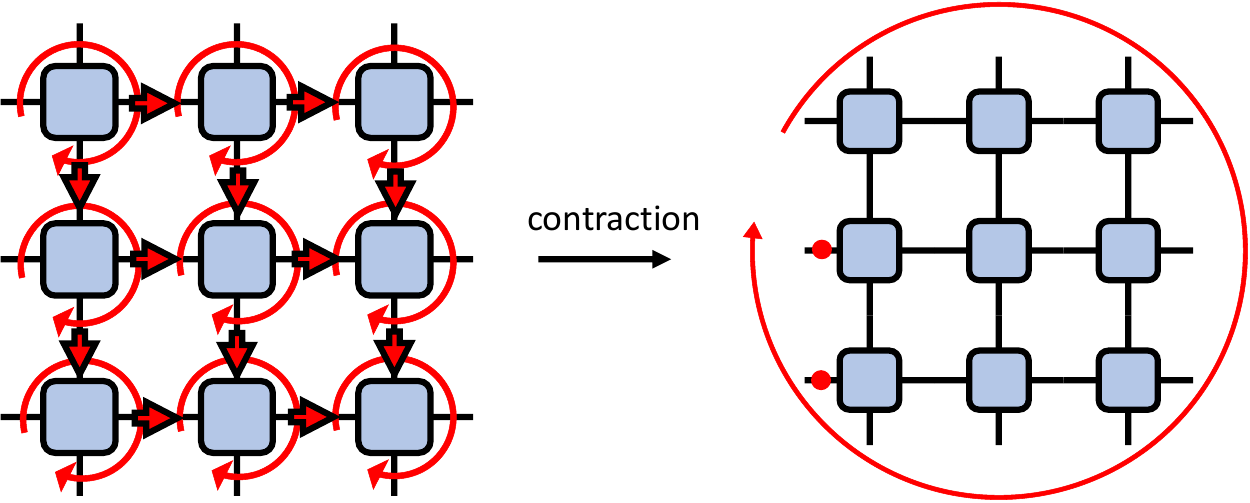}
	\caption{Contraction of fermionic tensors. The contracted tensor can be
	brought into standard ordering at the cost of introducing a sign-correction
	indicated by the red dots corresponding to matrices $s_{ij}=(-1)^{|j|}
	\delta_{ij}$.}
	\label{fig:rect}
\end{figure}

The assignments are as follows. We fix the orientation of all fermionic bonds to
be left to right for horizontal links and top to bottom for vertical links. We
define the index ordering of the tensors to be clockwise with the start point at
the top most index along the left edge (cf. Fig.~\ref{fig:rect}). We refer to this
pattern as \emph{standard ordering}. With this choice, the fermionized
contracted bosonic tensor network is identical to the contraction of the tensors
fermionized individually up to a correction of sign factors that depends solely
on the fermion parities of the fermionic modes at the left boundary of the
patch. 

\begin{figure*}
	\centering
\includegraphics[width=\textwidth]{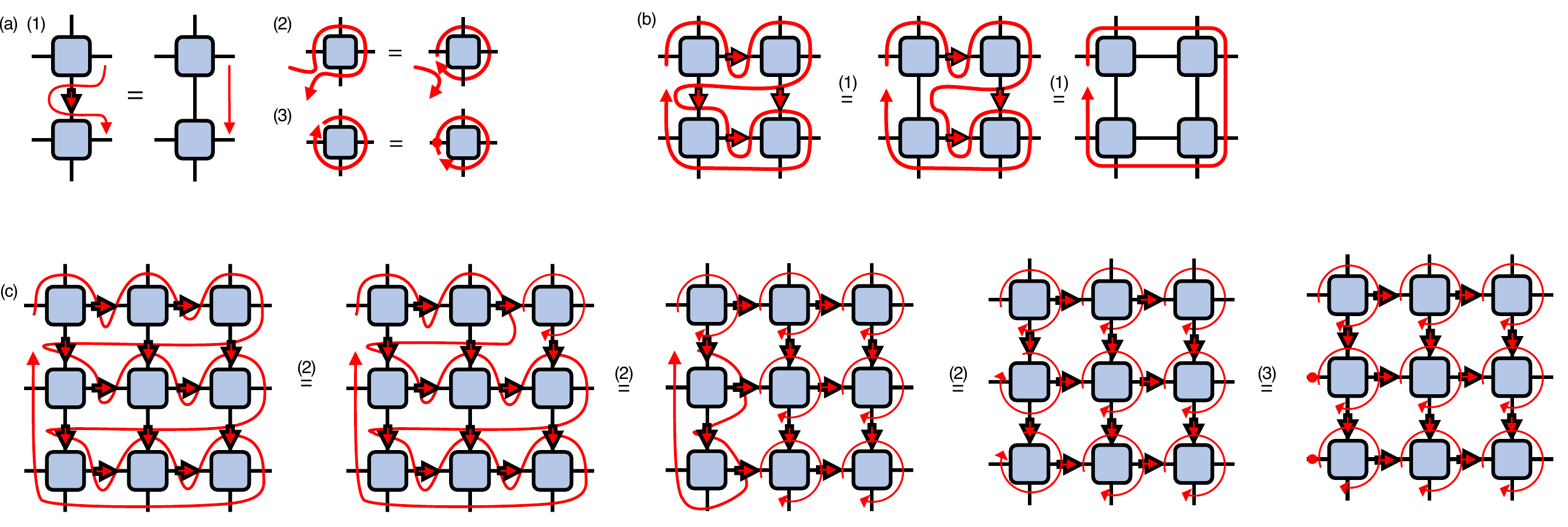}
	\caption{(a) Identities for contracting and re-ordering fermionic modes. (1) Contraction of a fermionic bond. 
 If the modes to be contracted
	(red arrow) occur in consecutive order in the ordering of all fermionic
	modes (red line) they can be integrated out and removed from the tensor
	network. (2) Manipulating the ordering of fermionic 
 modes. The total fermion
	parity of the tensors is even. Thus, when the ordering string `encircles' a
	tensor, we can decouple it from the ordering string. (3) Cyclic
	permutation of the fermionic modes of a parity even tensor according to
	$\theta_2^j \theta_3^k \theta_4^l \theta_1^i = (-1)^{i} \theta_1^i
	\theta_2^j \theta_3^k \theta_4^l$ yields a sign factor $(-1)^i$ represented as a red
	dot. (b) A tensor network with ordering $\mathcal O_\text{rect}$. Fermionic
	bonds are integrated out sequentially using rule (1) to arrive at a fully
	contracted network with boundary modes ordered in standard ordering. (c)
	Calculating the sign factor $\sigma(\mathcal O_\text{rect},\mathcal
	O_\text{initial})$. We deform the global ordering string according to rule
	(2) several times. In the last equality we use the cyclic permutation rule (3) to obtain
	the initial ordering. In this process we obtain the sign factor represented
	by the red dots along the left edge of the network. }
	\label{fig:contract}
\end{figure*}

To see this, we consider the initial ordering or all fermionic modes in the
uncontracted tensor networks $\mathcal O_\text{initial}$ and perform a
particular reordering to the ordering $\mathcal O_\text{rect}$. This produces a
sign factor $S=(-1)^{\sigma(\mathcal O_\text{initial},\mathcal O_\text{rect})}$.
We show that $\mathcal O_\text{rect}$ is compatible with the contraction of
fermionic bonds, i.e., we can readily perform the integration over Grassmann
variables and are left with a tensor that only has fermionic modes at its
boundary. In particular the ordering of the boundary modes coincides with the
standard ordering (see Fig.~\ref{fig:contract}(d)). We evaluate the sign factor
$S$ (see Fig.~\ref{fig:contract}(e) and find that it only depends on the
fermionic parities of the fermionic modes at the left boundary of the patch.

\subsection{Fermion to Boson mapping for correlation functions}
\label{app:bosonfermion2} 
\begin{figure}
	\centering
	\includegraphics[width=0.45\textwidth]{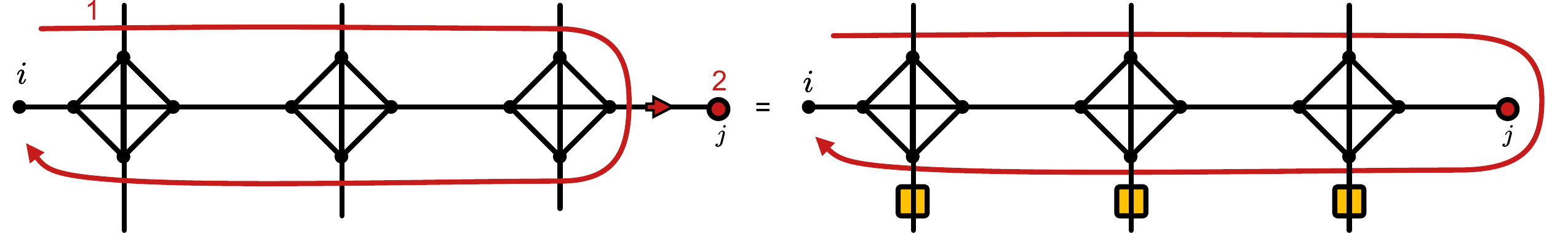}
	\caption{Region $A$ of the tensor network, containing the
	individual modes $\theta_i$ and $\theta_j$ and a string of connecting
	tensors such that the total parity of $A$ is even. Left: Division of the
	region $A$ into two parity odd patches (ordered). The first patch is contracted and emerges with standard ordering. The second one is given by the single mode
	$\theta_j$. Right: The contraction of the two
	patches emerges with non-standard ordering. Reordering it to standard-ordering brings about a string of sign factors along the modes at the bottom.}
		\label{fig:corr}
\end{figure}

We now describe the steps of the mapping in more
detail. On a square lattice (cf.~Fig.~\ref{fig:corr}), the product of a single
4-leg matchgate tensor and a pre-exponential $\theta$ factorizes into a 3-leg
matchgate tensor and a single $\theta$ as illustrated by the example
\begin{equation} 
e^{\frac 1 2 \theta_i a_{ij} \theta_j} \theta_3 = (1 + a_{12}
\theta_1 \theta_2 + a_{14} \theta_{1} \theta_4 + a_{24} \theta_2 \theta_4)
\theta_3 \;.
\end{equation}

We next divide the tensor network into the parts $A$ and $\bar A$ as indicated
above (cf.~Fig.~\ref{fig:corr_order}) and contract the tensor network $A$. Part
$A$ is the interesting part and is itself composed of fermion-parity even
tensors and two fermion-parity odd tensors whose ordering has to be stated
explicitly (as they do not commute). The ordering is inherited from the
definition of the correlation function, here it is $\theta_i \theta_j$. A
convenient partitioning of $A$ is to split-off $\theta_j$ from the rest.
Everything but $\theta_j$ can be contracted as usual and gives rise to a tensor
with fermionic modes in standard ordering. In our case, this is a clockwise
orientation with starting point on the top-most edge on the left side of the
patch. 

We now only need to contract the remaining $\theta_j$ mode with the rest and
reorder the product to standard ordering. To do so, we need to move $\theta_j$
past all fermionic modes such that standard ordering is reached. In the example
in Fig.~\ref{fig:corr_order} this corresponds to moving $\theta_j$ past all
modes on the bottom segment of the already contracted strip. This induces a sign
factor depending on the fermion parity of the corresponding edges as indicated
by yellow dots in the figure. With the standard ordering and even-parity in
place we can now transform the tensor network to a bosonic tensor network as
before and rewrite it as a weighted $\mathbb Z_2$-tensor network.

\section{Class D topological superconductors}
\label{sec:ClassDReview}

\subsection{Band structure, topological invariants, and Dirac representation}
\label{app:band}

The \emph{Bogoliubov-de Gennes Hamiltonian} of spin-triplet superconductor in
the absence of time reversal invariance (the defining signatures of symmetry
class D) is conventionally represented as 
\begin{align}
	\hat H:= (C^\dagger,C)\left( \begin{array}{cc}
	h & \Delta \\ 
	\Delta^\dagger & -h^T
	\end{array}\right)\left( \begin{matrix}
		C\cr C^\dagger
	\end{matrix} \right)=: 
    \psi^\dagger \tilde H \psi,
\end{align}   
where $C=\{C_i\}$ is a vector of (spinless) lattice fermion creation operators,
and $\Delta=-\Delta^T$ an anti-symmetric lattice order parameter. In this
language, the class D symmetry of the matrix Hamiltonian $\tilde{H}$ assumes the
form $H^T =-\sigma_x H \sigma_x$, the Pauli matrix acting in particle-hole
space. From here, we may pass to a real (`Majorana') representation via unitary
transformation 
\begin{align}
	\psi=\left( \begin{matrix}
		C\cr C^\dagger 	
	\end{matrix} \right):=\frac{1}{\sqrt{2}}\left( \begin{matrix}
		\eta+i \nu\cr \eta - i \nu
	\end{matrix} \right)=: 
    U \Theta,
\end{align} 
with $\Theta=(\eta,\nu)^T$, and $U=\frac{1}{\sqrt{2}}\left( \begin{smallmatrix}
	1&i\cr 1&-i \end{smallmatrix} \right)$. A quick calculation shows that in
	this \emph{Majorana representation} our Hamiltonian assumes the form
	$\hat{H}=i \Theta^T A \Theta$ with the real anti-symmetric matrix
\begin{align}
	\label{eq:MajoranaBdG}
	A=\left( \begin{matrix}
		-i h_A-i \Delta_+& h_S- \Delta_- \cr -h_S + \Delta_- & -i h_S +i  \Delta_+
	\end{matrix} \right),\\
	h_{S /A}=\frac{1}{2}(h \pm h^T),\qquad \Delta_\pm=\frac{1}{2}(\Delta  \pm \Delta^\dagger).
\end{align}
Conversely, any even-dimensional anti-symmetric matrix may be represented in the
block form Eq.~\eqref{eq:MajoranaBdG}, and in this way one may pass from the
complex Bogoliubov-de Gennes representation to the Majorana representation, and
back. 

According to the general classification of topological insulators and
superconductors, the two-dimensional superconductor in class D is a Chern
insulator, carrying a  $\Bbb{Z}$-valued  topological index $c$. This index is in
turn obtained by summation $c=\sum_{n=1}^N c_n$ over Chern numbers carried by
individual Bloch bands, $n=1,\dots,N$, with band energies below the
superconductor band gap at $\epsilon=0$. (Recall that single particle energies
of a superconductor occur in pairs $\pm \epsilon$, and that the presence of an
order parameter generically implies a band gap around $\epsilon=0$.) 

To compute the numbers $c_n$, we label the band eigenvectors as  $|n,k\rangle$
in terms of their two-dimensional crystal momenta $k\in \Bbb{T}^2$ and define
the \emph{Berry connection} 
\begin{equation}
a(n,k):=i \langle n,k |\dd| n,k\rangle = i \langle
n,k |\partial_{k^i} |n,k\rangle  \dd k^i 
\end{equation}
and its associated
\emph{curvature}
\begin{equation}
f(n,k):= \dd a(n,k)=i (\partial_{k^j}\langle n,k
|)(\partial_{k^i} |n,k\rangle) \dd k^j\wedge \dd k^i.
\end{equation}
Integration of the latter over the Brillouin zone then yields the \emph{Chern
numbers} as  
\begin{equation}
c_n:= \frac{1}{2\pi}\int_{\Bbb{T^2}} f(n,k). 
\end{equation}
One is often interested in detecting \emph{changes} of topological invariants
upon crossing a topological phase transition point, i.e., situations where the
Chern numbers $(c_n,c_{n+1})\to (c_n \pm 1,c_{n+1}\mp 1)$ of two neighboring
bands change via the transient appearance of a band gap closure at a Dirac
point. The simplest possible case  $(c_n,c_{n+1})=(0,0)\to (\pm 1,\mp 1)$ may be
conveniently described in terms of a minimal two-band reduction, where the
Hamiltonian reduced to the two bands is described as a two-dimensional matrix
Hamiltonian, $H_k^{(2)}:= \sum_{a=1}^{3} c_a(k)\sigma_a$, and the class D
symmetry in Majorana representation 
\begin{equation}
(H^T)_k=H^T_{-k}=- H_k 
\end{equation}
requires $c_1(-k)=-c_1(k), c_2(k)=c_2(k),c_3(-k)=-c_3(k)$. In the vicinity of a
phase transition point, the Hamiltonian reduces to the Hamiltonian
Eq.~\eqref{eq:DiracHamiltonian}. In this reduction, the above definition of
Chern numbers counts the number of windings of the coefficient vector
$(q_1,q_2,m+c q^2)$ around the origin as a function of $q$. It is
straightforward to verify that this criterion translates to the assignment of
Chern numbers mentioned in the main text.

\subsection{Majorana zero mode} \label{app:majorana}

We may upgrade our band insulator to one with a topologically degenerate ground
space by adding $\Bbb{Z}_2$ vortices. Assuming a gauge where the entire
$\pi$-flux carried by a half flux quantum vortex is picked up along one lattice
bond, a $\Bbb{Z}_2$ vortex is realized by the insertion of a line defect along
which the sign of the hopping amplitude is inverted, $1 \to -1$. On the same
basis, a vortex pair corresponds to a line defect of finite length. Now assume
each of its two vortex centers to be surrounded by a region with $a>a_+$ in
which the system is topologically trivial, or represents an effective vacuum.
One may  show (see Appendix \ref{app:majorana}) that at the boundaries  to the
topologically non-trivial outer region with $a<a_+$  zero energy Majorana
states, $\gamma$, forms. (While a finite vortex core region facilitates the
identification of the edge mode, its size may be shrunk to zero without
compromising the topologically protected Majorana state.) The two Majoranas
bound by a vortex pair define a complex fermion state, $\psi$,  whose occupation
parity realizes the $\Bbb{Z}_2$ degeneracy of the bi-vortex ground state. While
the vortices  cannot be individually removed, one may `fuse' them either by
contracting the connecting string, or by extending the vortex disk areas.
Depending on the occupation of the previously non-local fermion, one is then
either left with a vacuum state, or with a local and gapped fermion, in
accordance with the \emph{fusion rule} $\gamma\otimes \gamma=1\oplus \psi$.       

\begin{figure}[h]
	\centering
	\includegraphics[width=0.95\linewidth]{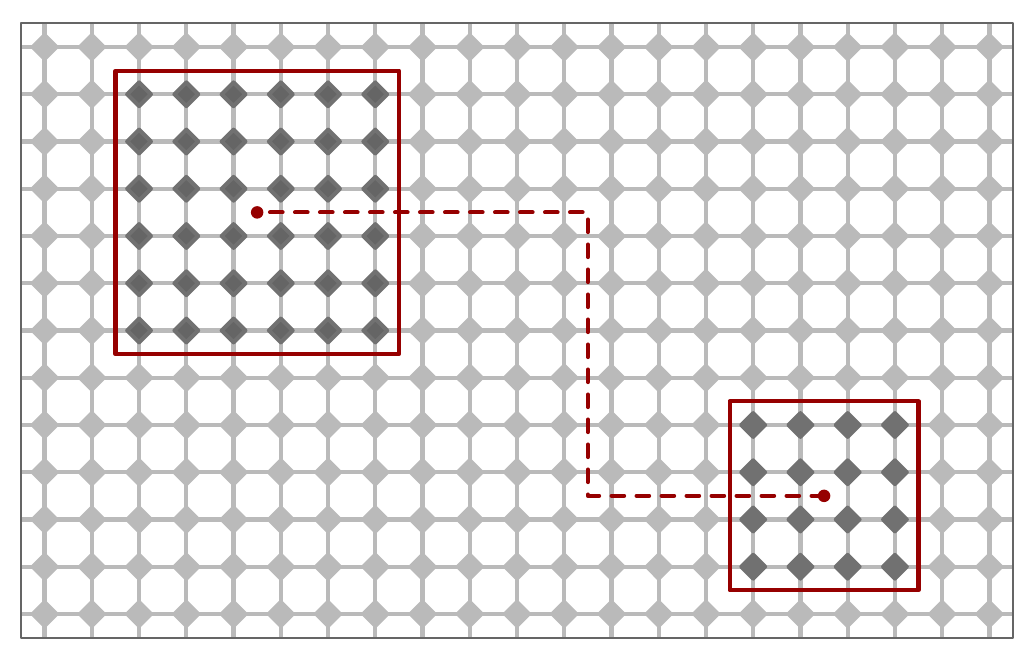}
	\caption{A pair of $\Bbb{Z}_2$ vortices realized by a string of sign
	inverted hopping amplitudes (bonds intersected by dashed line). At the
	interface between regions with $a>a_+$ (dark shaded) and $a<a_+$ (light
	shaded) zero energy Majorana edge modes form. The fusion of the two vortices
	leads to a topologically trivial configuration (equivalently, a
	$2\pi$-vortex)}   
	\label{fig:VortexPair}
\end{figure}

\begin{figure*}
	\centering
	\includegraphics[width=\textwidth]{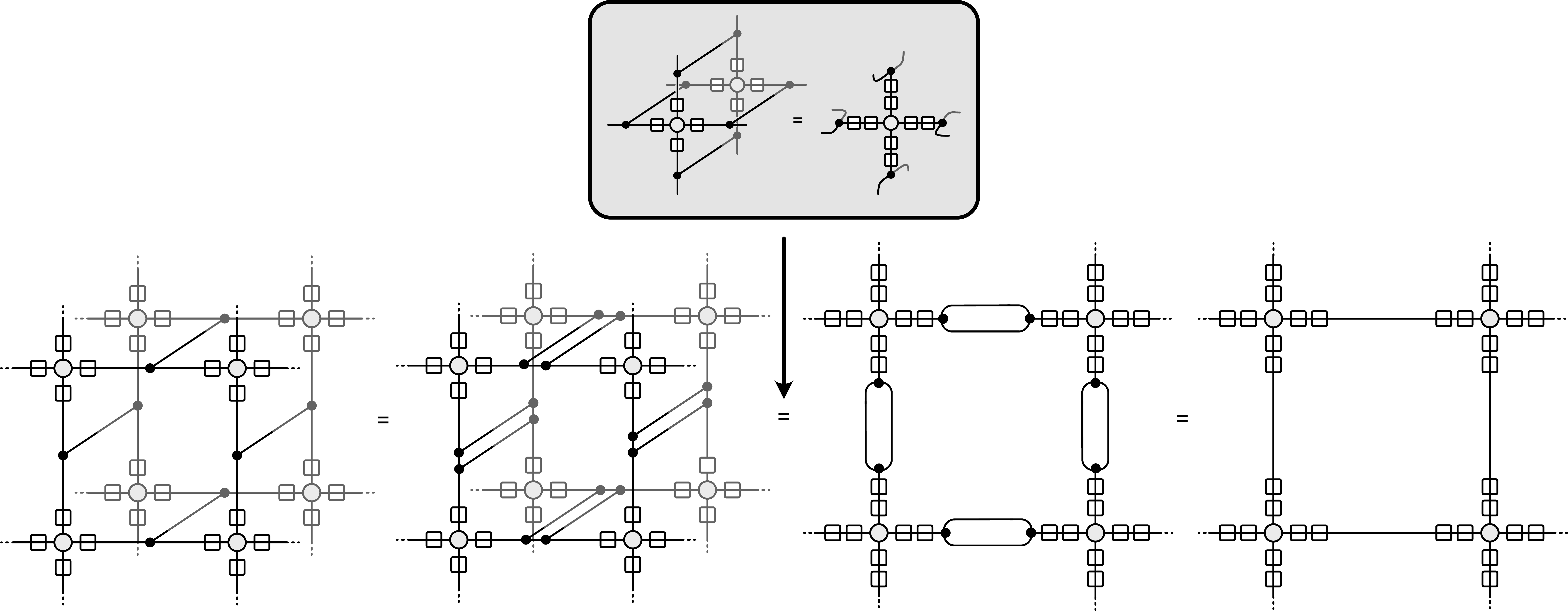}
	\caption{Collapsing the double layer $\braket{\Psi(t)|\Psi(t)}$ TN to a
	single layer TN via a sequence of local deformations. All identities used
	follow from the definitions of the $\mathbb Z_2$- and the $\delta$-tensor in
	Fig.~\ref{fig:transform}(b)}
	\label{fig:tc_details}
\end{figure*}

\paragraph*{Vortex zero mode.} For completeness, we here demonstrate the
formation of the vortex state and its zero mode by explicit computation.
Assuming a core sufficiently large to admit a continuum representation in terms
of the effective Hamiltonian Eq.~\eqref{fig:Dispersion}, we consider a disk of
radius $R$  with topological control parameter $m<0$ and $m>0$ inside and
outside the disk perimeter.  With $k_i = -i \partial_i$, and switching to polar
coordinates we have $\partial_1=\cos \phi \partial_r - \frac{1}{r}\sin \phi
\partial_\phi$ and $\partial_2=\sin \phi \partial_r + \frac{1}{r}\cos \phi
\partial_\phi$. Using these identities,   the Dirac Hamiltonian (with $\kappa=1$
for simplicity) assumes the form 
\begin{align}
  H &= i(\cos \phi \sigma_1 + \sin \phi \sigma_3)\partial_r + \cr 
  &\qquad \frac{i}{r}\left( - \sin \phi \sigma_1 + \cos \phi \sigma_3  \right)\partial_\phi + m(r) \sigma_2. 
\end{align}
We are interested in the existence of zero mode solutions $H \psi =0$. The
structure of the Hamiltonian suggests that it should be possible to apply a
rotation around $\sigma_2$ to bring it to a simpler form. Specifically, consider
the equivalent equation $\tilde{H}\tilde{\psi}=0$, with $\tilde{H}= U H U^{-1}$
and $\tilde{\psi}=U \psi$ where $U=\exp(-\frac{i}{2}\phi \sigma_2)$. It is
straightforward to verify that 
\begin{align*}
  \tilde{H}= i \left(\partial_r + \frac{1}{2r}\right) \sigma_1 + \frac{i}{r}\partial_\phi \sigma_3 + m(r) \sigma_2 .
\end{align*}   
(Note: Naively, the second term in parentheses appears to violate Hermiticity.
However, this is not so. It is actually required to make $i
\left(\partial_r+\frac{1}{2r} \right) $ Hermitian w.r.t.~the measure $rdr$ of
polar integration.) Assuming $\phi$-independence of the zero mode, we are left
with the equation
\begin{align*}
  \left( \partial_r + \frac{1}{2r}  - m(r) \sigma_3 \right)\tilde{\psi}(r)=0.
\end{align*}
Now decompose $|\tilde{\psi}\rangle =\psi_+|\uparrow\rangle + \psi_-
|\downarrow\rangle$ in terms of eigenvectors of  $\sigma_3 $. This leads to 
\begin{align*}
  \left( \partial_r + \frac{1}{2r}  \mp m(r) \sigma_3 \right)\psi_\pm(r)=0,
\end{align*}
with the solutions (pre-normalization)
\begin{align*}
  \psi_\pm(r) = \left(\frac{R}{r}\right)^{1 /2}\exp \left(\pm \int_R^r ds\, m(s) \right).
\end{align*}
Since $m(r)>0$ for $r>R$, we retain $\psi_-$ as a normalizable option. Tidying
up, we obtain
\begin{align*}
  \psi(r,\phi)=\mathcal{N} \left(\frac{R}{r}\right)^{1 /2}\exp \left(- \int_R^r ds\, m(s) \right) e^{i \frac{\phi}{2}}|\downarrow\rangle 
\end{align*}  
for the zero mode of our system. At this point, it becomes clear why a (half-)
vortex is required to realize this solution: the wave function as such obeys
anti-periodic boundary conditions, $\psi(r,\phi+2\pi)=-\psi(r,\phi)$. Such
states are not tolerated. However, if we assume  $\pi$-flux inserted at the
origin, the additional phase twist adds to that of $\tilde \phi$ to define a
state with periodic boundary conditions. In the Majorana representation with its
real anti-symmetric Hamiltonian, the $\pi$-flux insertion may be realized in
terms of a line defect of bonds with sign inverted  hopping amplitude as
indicated in Fig.~\ref{fig:VortexPair}.

\section{Details on tensor network contractions for toric code ground states}
\label{app:tc}

\begin{figure}[h]
	\centering
	\includegraphics[width=0.45\textwidth]{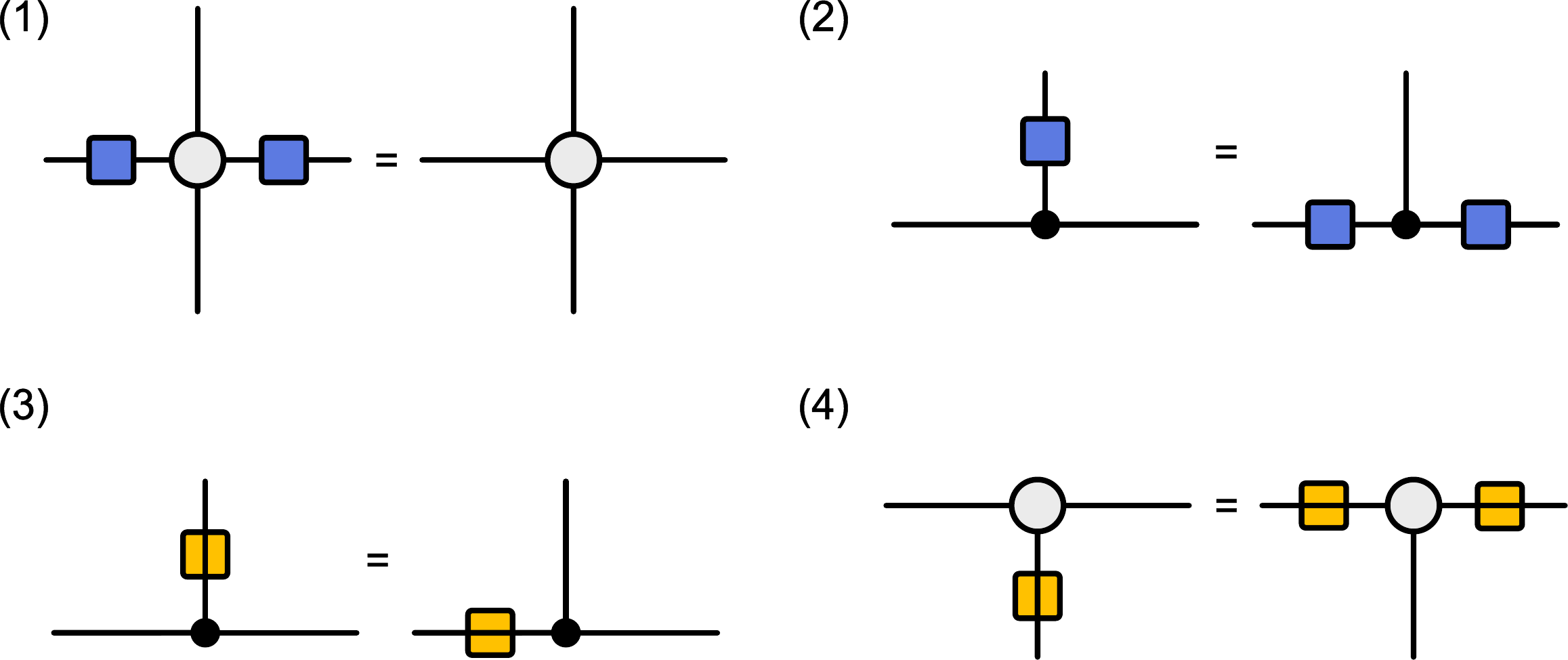}
\caption{TN identities for the interaction of Pauli matrices $\sigma_x$ (blue
	square) and $\sigma_z$ (yellow square) with $\mathbb Z_2$-tensors (grey
	circles) and $\delta$-tensors (black circles). (1) A pair of
	$\sigma_x$-matrices applied to any pair of indices can be absorbed by a $\mathbb
	Z_2$-tensor. (2) The same holds for $\sigma_z$-matrices and $\delta$-tensors which is equivalent to identity (3). A
	string of $\sigma_x$($\sigma_z$)-matrices can be pulled through a
	$\delta$($\mathbb Z_2$)-tensor as shown in identities (2) and (4).  }
	\label{fig:z2_id}
\end{figure}

\begin{figure*}[h]
	\centering
	\includegraphics[width=\textwidth]{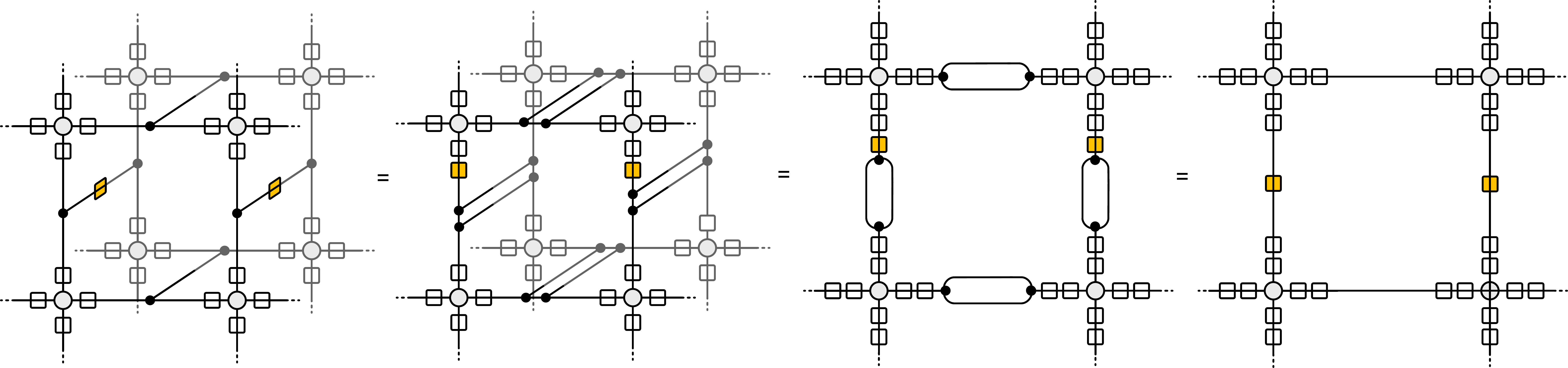}
\caption{The overlap $\braket{\Psi(t)|S_z|\Psi(t)}$ reduced to a single layer
MGTN with sign-string following the same reductions as in
Fig.~\ref{fig:tc_details}.}
\label{fig:app_string}
\end{figure*}

\begin{figure*}[h]
	\centering
	\includegraphics[width=\textwidth]{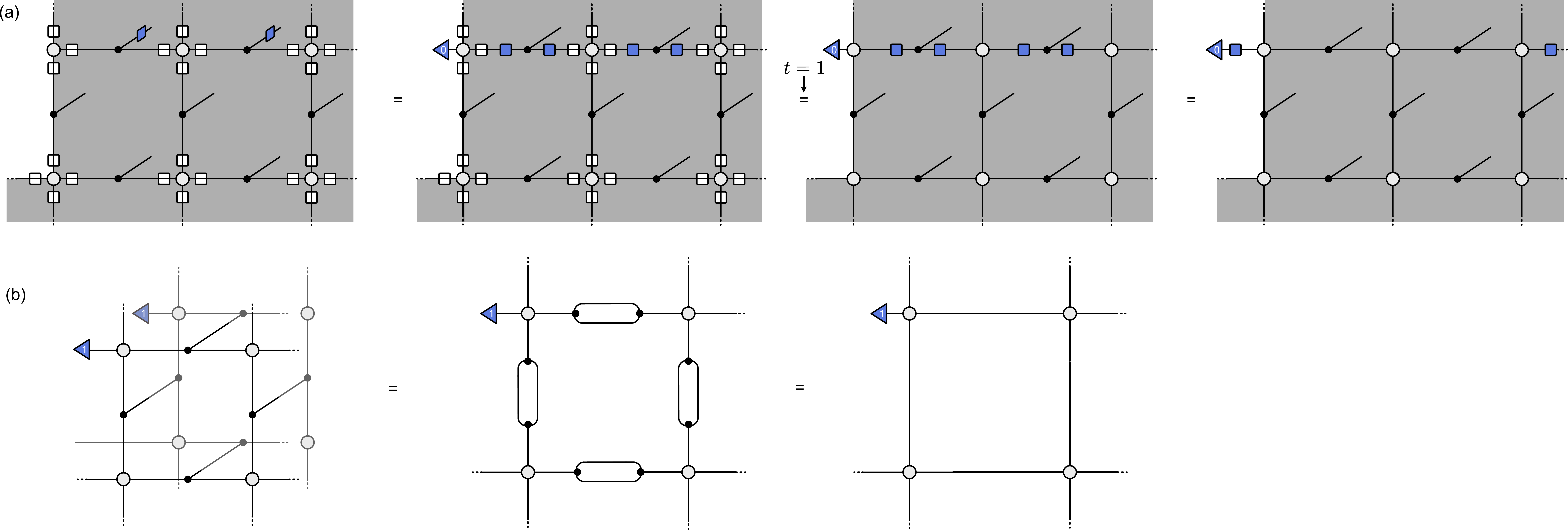}
\caption{(a) Local patch of the state vector  $S_x\ket{\Psi_\square}$. In the second
equality we restrict to the fix-point, $t=1$. For $t\neq 1$, the
non-commutativity of $\sigma_x$ and the weight-matrices obstructs the simple
reduction procedure. (b) Contracting $\braket{S_x \Psi_\square|S_x
\Psi_\square}$ to a single layer TN.}
\label{fig:app_sx}
\end{figure*}

In this appendix, we provide some details on tensor network contractions and
identities used to derive results for the toric code ground state expectation
values. The identity in Fig.~\ref{fig:tc}(b) is derived using a sequence of
local deformations shown in Fig.~\ref{fig:tc_details}. Here, we use that
$\delta$-tensors can be duplicated and that a local double layer patch can be
reduced to a single layer patch as shown in the inset of the figure. Finally, we
use that a loop of $\delta$-tensors can be contracted to an identity matrix as
shown in the last equality.

Next, we provide further details on the interpretation of the TN representing a
correlation function in Sec.~\ref{subsec:tc} Fig.~\ref{fig:corr_order} (center)
as a ground state expectation value of the toric code in the case, where the
parameter $a$ determining the weight matrices in the tensor network are $a=1$
inside and $a=0$ outside a region $R$.

In particular we show how a TN with a sign-string can be interpreted as
$\braket{\Psi_\square |S_z|\Psi_\square}$ and how the TN with parity violations
at sites $i$ and $j$ originating from the $\ket{1}$-state projections is given
by $\braket{S_x \Psi_\square |S_x \Psi_\square}$. All of these identifications
are derived using the general rules for the `interaction' of $\sigma_x$ and
$\sigma_z$ matrices with $\mathbb Z_2$- and $\delta$-tensors shown in
Fig.~\ref{fig:z2_id}.

\subsection{Sign string}\label{app:string} Here, we show how the expectation
value $\braket{\Psi_\square |S_z|\Psi_\square}$ is identified with a TN with a
string of sign-flips (generated by $\sigma_z$-matrices) on the dual lattice. To
see this consider Fig.~\ref{fig:app_string}. On the very left, we show a local
patch of the double layer of $\ket{\Psi_\square}$ and $\bar{\Psi_\square}$
sandwiching a physical $\sigma_z$-error string. We again use a $\delta$-tensor
doubling and then make use of the fact, that a $\sigma_z$ matrix can be moved
through a $\delta$-tensor according to identity (3) in Fig.~\ref{fig:z2_id}.
After the contraction of the $\delta$-tensor loops, we obtain a purely virtual
TN with the sign flip string as desired.

\subsection{$\ket{1}$-projections and parity violations}\label{app:parity} In
the main text, we have shown how the $\ket{1}$-projections can be gauged away.
Here, we provide an alternative interpretation and show how the projections onto
the $\ket{1}$-state can be interpreted as the overlap $\braket{S_x
\Psi_\square|S_x \Psi_\square}$ establishing a connection with the
$m$-excitations in the toric code. We start by taking a closer look at the state
vector $S_x \ket{\Psi_\square}$ and its tensor network representation in
Fig.~\ref{fig:app_sx}(a). By using identity (2) from Fig.~\ref{fig:z2_id} we can
trade a $\sigma_x$-matrix acting on the physical index for two
$\sigma_x$-matrices acting on the virtual level. We also note that a 3-leg
$\mathbb Z_2$-tensor is identical to a 4-leg $\mathbb Z_2$-tensor with one index
projected to the state vector $\ket{0}$ (first equality in
Fig.~\ref{fig:app_sx}(a)). Now, if we are working with a TC ground state with
string tension ($a\neq 1$), we encounter the problem that the
$\sigma_x$-matrices do not commute with the weight-matrices which obstructs
further simplifications. However, at the fix-point ($a=1$), the weight matrices
reduce to identity matrices (second equality in Fig.~\ref{fig:app_sx}(a)). In
this case, we can absorb pairs of $\sigma_x$-matrices at the $\mathbb
Z_2$-tensors using identity (1) from Fig.~\ref{fig:z2_id}. We are left with
$\sigma_x$-matrices acting on the $\ket{0}$-states at the end of the string
$S_x$ which give rise to the $\ket{1}$-state projections we set out to identify.

Now, we consider the overlap $\braket{S_x \Psi |S_x \Psi}$ and show that it
gives rise to a single layer TN with $\ket{1}$-projections at the end of the
$S_x$-string. A local patch of the double layer TN around site $i$ is shown in
Fig.~\ref{fig:app_sx}(b). In analogy to the reductions in
Fig.~\ref{fig:tc_details}, we can reduce the TN to a
single layer TN with $\ket{1}$-projections at the site $i$.

\end{document}